\documentclass{aa}
\usepackage[varg]{txfonts}
\usepackage{natbib}
\bibpunct{(}{)}{;}{a}{}{,} 
\usepackage{graphicx}
\usepackage{hyperref}
\usepackage{booktabs}
\usepackage{color}
\usepackage{soul}
\usepackage{ulem}
\usepackage{amsmath}
\usepackage{float}

\begin{document}

\title{Newly formed downflow lanes in exploding granules in the solar photosphere}

\author{M. Ellwarth\inst{\ref{KIS},\ref{IFA}}
        \and C. E. Fischer\inst{\ref{KIS}}
        \and N. Vitas\inst{\ref{IAC},\ref{UNITEN}}
        \and S. Schmiz\inst{\ref{KIS}}
        \and W. Schmidt\inst{\ref{KIS}}}

\institute{Leibniz Institute for Solar Physics (KIS), Sch\"oneckstrasse 6, 79104 Freiburg, Germany\label{KIS} \and Georg-August Universität Göttingen, Institut für Astrophysik, Friedrich-Hund-Platz 1, 37077 Göttingen, Germany\label{IFA} \and Instituto de Astrof\'{i}sica de Canarias, 38205 La Laguna, Tenerife, Spain\label{IAC} \and Departamento de Astrof\'isica, Universidad de La Laguna, 38205 La Laguna, Tenerife, Spain\label{UNITEN}}

\abstract { Exploding granules have drawn renewed interest because of their interaction with the magnetic field (either emerging or already present). Especially the newly forming downflow lanes developing in their centre seem to be eligible candidates for the intensification of magnetic fields. We analyse spectroscopic data from two different instruments in order to study the intricate velocity pattern within the newly forming downflow lanes in detail.}
{ We aim to examine general properties of a number of exploding granules, such as their lifetime and extend. To gain a better understanding of the formation process of the developing intergranular lane in exploding granules, we study the temporal evolution and height dependence of the line-of-sight velocities at their formation location. Additionally, we search for evidence that exploding granules act as acoustic sources.}
{ We investigated the evolution of several exploding granules using data taken with the Interferometric Bidimensional Spectrometer and the Imaging Magnetograph eXperiment. Velocities for different heights of the solar atmosphere were determined by computing bisectors of the Fe I 6173.0\,\AA~and the Fe I 5250.2\,\AA~lines. We performed a wavelet analysis to study the intensity and velocity oscillations within and around exploding granules. We also compared our observational findings with predictions of numerical simulations.} 
{Exploding granules have significantly longer lifetimes (10 to 15 minutes) than regular granules. Exploding granules larger than 3.8$\,\arcsec$ form an independent intergranular lane during their decay phase, while smaller granules usually fade away or disappear into the intergranular area  (we find only one exception of a smaller exploding granule that also forms an intergranular lane). For all exploding granules that form a new intergranular downflow lane, we find a temporal height-dependent shift with respect to the maximum of the downflow velocity. Our suggestion that this results from a complex atmospheric structure within the newly forming downflow lane is supported by the comparison with synthesised profiles inferred from the simulations.  We found an enhanced wavelet power with periods between 120\,s to 190\,s seen in the intensity and velocity oscillations of high photospheric or chromospheric spectral lines in the region of the dark core of an exploding granule.}
{}
\date{Received 24 April 2020 / Accepted 21 June 2021}

\maketitle 

\section{Introduction}
\vspace{0.18cm}
Observations of the solar photosphere show a granular pattern that is driven by convection, with granules evolving on timescales of minutes and varying in shapes and sizes. An easily identifiable group of granules are the so-called exploding granules. They can be distinguished from regular granules by their rapid horizontal expansion with speeds between 1.7\,km s$^{-1}$ and 3.2\,km s$^{-1}$ \citep{1986A&A...161...31N}, and especially by the dark core they develop, which is seen as a reduction in continuum intensity. Exploding granules were first described by \cite{1968CRASB.266..199C}. They found that exploding granules show a specific way of dissolving. After developing a dark core in their centre, which is surrounded by a bright doughnut-shaped structure, these granules split into a new generation of granules. \cite{1986A&A...161...31N} found that exploding granules occupy about 2.5\% of the solar surface at any time and can reach diameters up to $5.5\arcsec$. In comparison, regular granules have a typical size of about $1.8\arcsec$ \citep{1986SoPh..107...11R}. 
In a recent study by~\cite{2020A&A...641A..50R}, a wide range of instruments were used to study the evolution of hundreds of exploding granules in continuum time sequences at differing spatial resolution. The authors showed that the highest expansion velocities are reached initially, followed by a rapid decrease in the expansion velocity within the first two minutes.

First observations of exploding granules in white-light images caused authors to speculate about the origin of the intensity drop in the core. Based on their models of solar granulation, \cite{1978ApJ...222L..69N} found that granules twice the size of a mean granule show a darker centre surrounded by a bright ring, resembling an exploding granule. In the model, this was caused by opacity fluctuations in large-scale granules. \cite{2001A&A...368..652R} used spectrographic data taken at different wavelengths to describe the temporal evolution of intensities and Doppler velocities of exploding granules. In most cases, they still measured an upflow velocity within the dark core, but these velocities became decreased in comparison of the surrounding gas. \cite{2001A&A...367.1011H} in contrast found a downflow velocity within the dark core. They used data from high-resolution 2D spectroscopic observations with the German Vacuum Tower Telescope at the Observatorio del Teide (Tenerife) and determined a downflow of material at the dark centre with a maximum velocity of about 0.7 to 1.2 km s$^{-1}$. This downflow was also confirmed by \cite{Nesis2001}. In the further development of exploding granules, the dark centre most often develops into new intergranular lanes, splitting the large granule into several fragments.
\cite{2012A&A...537A..21P} used data from the Imaging Magnetograph eXperiment (IMaX) on board the \textsc{Sunrise} stratospheric telescope \citep{2011SoPh..268...57M,2011SoPh..268....1B} to observe mesogranular-sized exploding granules and found magnetic loops embedded in the new downflow lanes of one of the fragmented exploding granules.

\cite{2017A&A...602L..12F} used spectropolarimetric data from Hinode and slit spectra from the Interface Region Imaging Spectrograph (IRIS) to observe the evolution of an exploding granule. They found that the magnetic elements at the edges of the observed granule are squeezed by fast horizontal flows and concluded that exploding granules can create shock waves in this way that dissipate in the chromosphere.

The phenomenon of exploding granules was also studied in numerical simulations of the photosphere \citep{1990ARA&A..28..263S,1993ApJ...408L..53R,1995ApJ...443..863R,2018ApJ...859..161R}. \cite{1995ApJ...443..863R} studied the formation of exploding granules and the temporal evolution of horizontal and vertical velocities with simulations. They found that the highest vertical upflows are located at the edge of a granule, with the downflow developing preferentially at the centre of the rapidly expanding granule. 

The turbulent atmosphere of the newly forming downflow lane could be an additional source for an increased acoustic flux. \cite{2010ApJ...723L.175R} found that exploding granules contribute to the excitation of solar p-modes, similar to the excitation of p-modes triggered in intergranular lanes \citep[see][]{2010ApJ...723L.134B}.

\cite{1995ApJ...444L.119R} and \cite{1998ApJ...495L..27G} used observations carried out with the Vacuum Tower Telescope, also known as the Dunn Solar Telescope (DST), of the National Solar Observatory at Sacramento Peak, New Mexico, and found that acoustic events occur primarily in intergranular lanes. \cite{1995ApJ...444L.119R} also observed several exploding granules, but found no significant correlation between acoustic events and exploding granules. On the other hand, \cite{Rutten2008} found that an exploding granule caused an increase in oscillation amplitude in the H$\alpha$ core intensity and H$\alpha$ Dopplergrams in one case study. 

Recent studies such as \cite{2020ApJ...896...62G} explored the interaction of the exploding granules with the emerging magnetic field. They linked the modification of the granule pattern to the evolution of the magnetic flux emergence. Analysis of simulations by \cite{2018ApJ...859..161R} also recently revealed the amplification of magnetic fields in the newly forming downflow lanes of exploding granules. These studies further prompt us to revisit the study of the characteristics of exploding granules and to determine the velocity field within the newly forming downflow lanes by now making use of currently available high-quality spectroscopic data.
We focus on the development of the dark centre of exploding granules and its formation into a new downflow lane. We study the general evolution as it appears in the continuum images and the Doppler velocities throughout the photosphere using a bisector analysis. We also investigate if there is evidence of an increased acoustic activity associated with the newly forming dark lane. Additionally, we use simulations  of the solar atmosphere from the \textsc{M\tiny{ANCHA}}3D code \citep{2017A&A...604A..66K} to compare our outcomes from the observations with current atmospheric models.

\section{Observations and data processing}
For our analysis of exploding granules, we used spectroscopic data close to the disc centre of the quiet Sun. In this section we give an overview on the scientific instruments, the observational settings, and the method we used to calculate the bisector Doppler velocities.

The data sets consist of time series recorded by IBIS \citep*{2006SoPh..236..415C}, which was mounted at the ground-based DST and the IMaX instrument \citep{2011SoPh..268...57M} on board the balloon-borne telescope \textsc{Sunrise} \citep{2011SoPh..268....1B}. Table \ref{tab:values} summarises the observation setup for both instruments.

\begin{table}
        \centering
        \small
        \begin{tabular}{llll}
                \toprule
                \toprule
                & &IBIS & IMaX \\
                \midrule
                $\lambda$ & [\AA]&{\bf Fe I \,\,\,\,6173.0} & {\bf Fe I \,\,\,\,5250.2 } \\
                $\Delta \lambda$&[\AA]&0.030 & 0.035\\
                Cadence&[s]&49 \tiny{(12th Oct)}& 29 \\
                &&31 \tiny{(14th Oct)}&\\
                \midrule
                $\lambda$ & [\AA]&{\bf Na I \,\,\,5896.0} &  \\
                $\lambda$ range &[\AA]&[-0.98:+0.83] &non-equidist. sampling\\
                Cadence&[s]&31 &\\
                \midrule
                $\lambda$ & [\AA]&{\bf Ca II 8542.0 } & \\
                $\lambda$ range&[\AA]&[-2.23:+2.47]& non-equidist. sampling\\
                Cadence&[s]&49 &\\
                \midrule
                R&&100000&75000  \\
                CCD pixel size&[arcsec]&$0.098\times 0.098$&$0.055\times 0.055$\\
                Field-of-View & [arcsec]&$38\times 86$&$ 51\times 51$\\
                Spatial res.&[arcsec]&$\sim 0.2$&$\sim 0.15$ \\
                Datatype&&Full Stokes&Stokes I \& V\\
                \bottomrule
        \end{tabular}
        \caption{Setup information for the IBIS and IMaX observations.}
        \label{tab:values}
\end{table}
\subsection{IBIS observations}
The data were taken in a quiet-Sun region close to the disc centre on 12 and 14 October 2016 during the service mode campaign of the DST. The duration of the time sequence in both cases was about 75 minutes. The IBIS data we employed consist of the spectropolarimetric measurements of the photospheric Fe I 6173.0\,\AA~line with equidistant wavelength steps of $0.030\,$\AA~as well as the chromospheric lines of Na I 5896.0\,\AA~and Ca II 8542.0\,\AA. The data were calibrated using a calibration package for IBIS data, which is offered by the National Solar Observatory (NSO). The calibration code includes the dark and flatfield correction and a routine to eliminate time-dependent image distortions ("de-stretch"). The field-dependent wavelength shift, which is due to the collimated mount of the etalons, is removed, as are the intensity gradients introduced by the narrowband prefilter.
A description of the full data set acquired on these days can be found in \cite{Mthesis}.

\subsection{IMaX observations}
The IMaX instrument is an imaging vector magnetograph that performed spectropolarimetric measurements in the Fe I 5250.2\,\AA~line. The data were taken on 10 June 2009 during the flight of the balloon-borne telescope \textsc{Sunrise}. It observed the solar atmosphere without atmospheric disturbances, leading to a high spatial resolution of 0.15\arcsec - 0.18\arcsec. IMaX has several observing modes. We used data obtained in the L12-2 observing mode, which is a longitudinal mode in which the Stokes parameters I and V were measured at 12 wavelengths. The step size was $0.035$\,\AA~between adjacent scan positions from -192.5 to +192.5\,m\AA~around the line core at Fe I 5250.2\,\AA. For the analysis we used the science-ready data set provided by the instrument team \citep{2011SoPh..268...57M}.

\begin{figure}
        \includegraphics[width=0.46\textwidth]{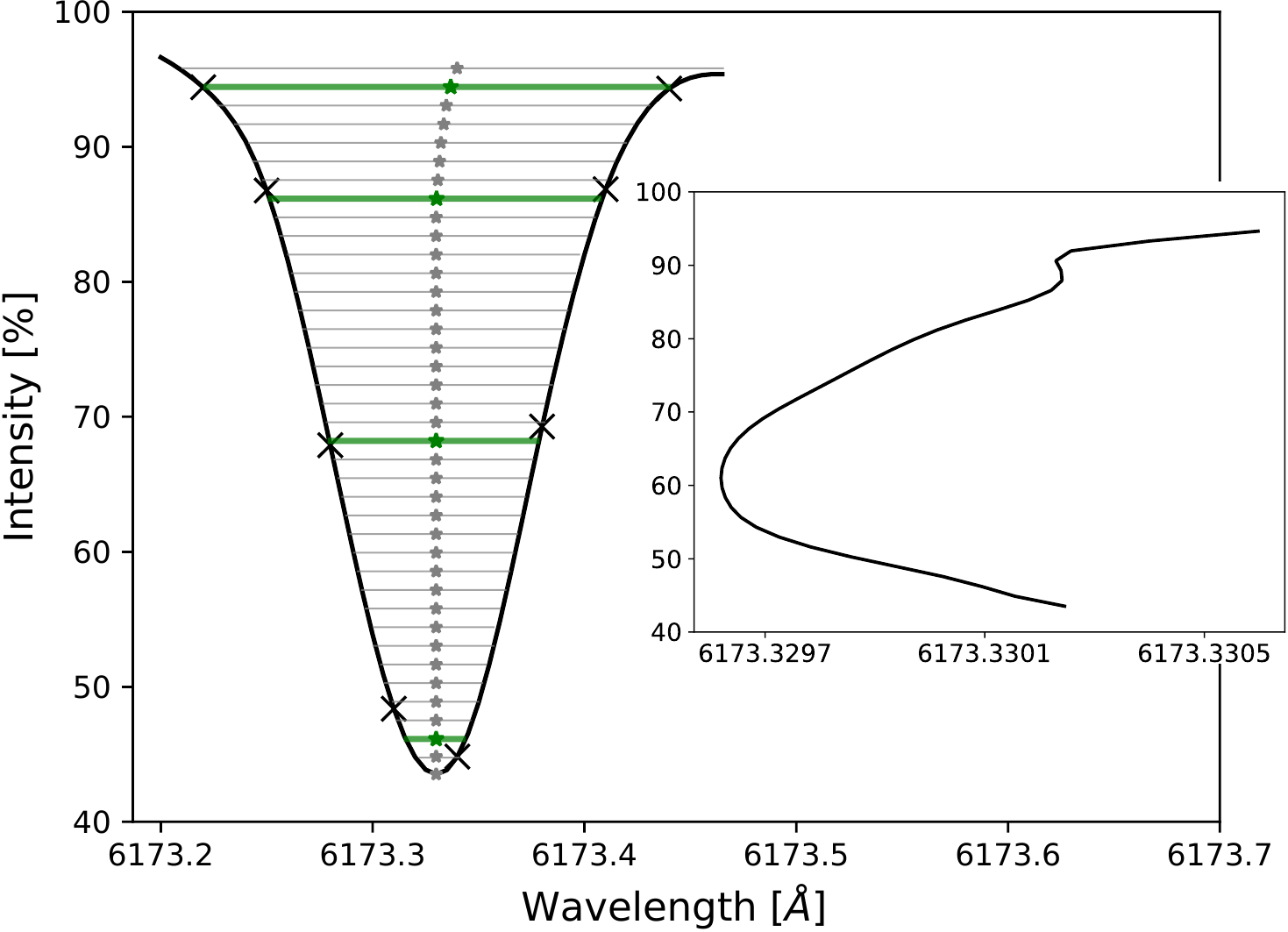}
        \caption{Line profile of the 6173.0\,\AA~line (IBIS data), where 100\% of intensity is designated as an average of the continuum. The crosses denote the spatially and temporally averaged observed line profile, and the solid line corresponds to a cubic spline interpolated profile. The horizontal lines show the calculated bisector levels, where the green lines mark the bisectors that are closest to the observed data points. The asterisk curve illustrates the bisector of the line profile. The smaller image on the right shows the bisector line resolved with an expanded wavelength scale.}
        \label{fig:bisec}
\end{figure}

\subsection{Bisector calculation}

\begin{figure}
        \includegraphics[width=0.44\textwidth]{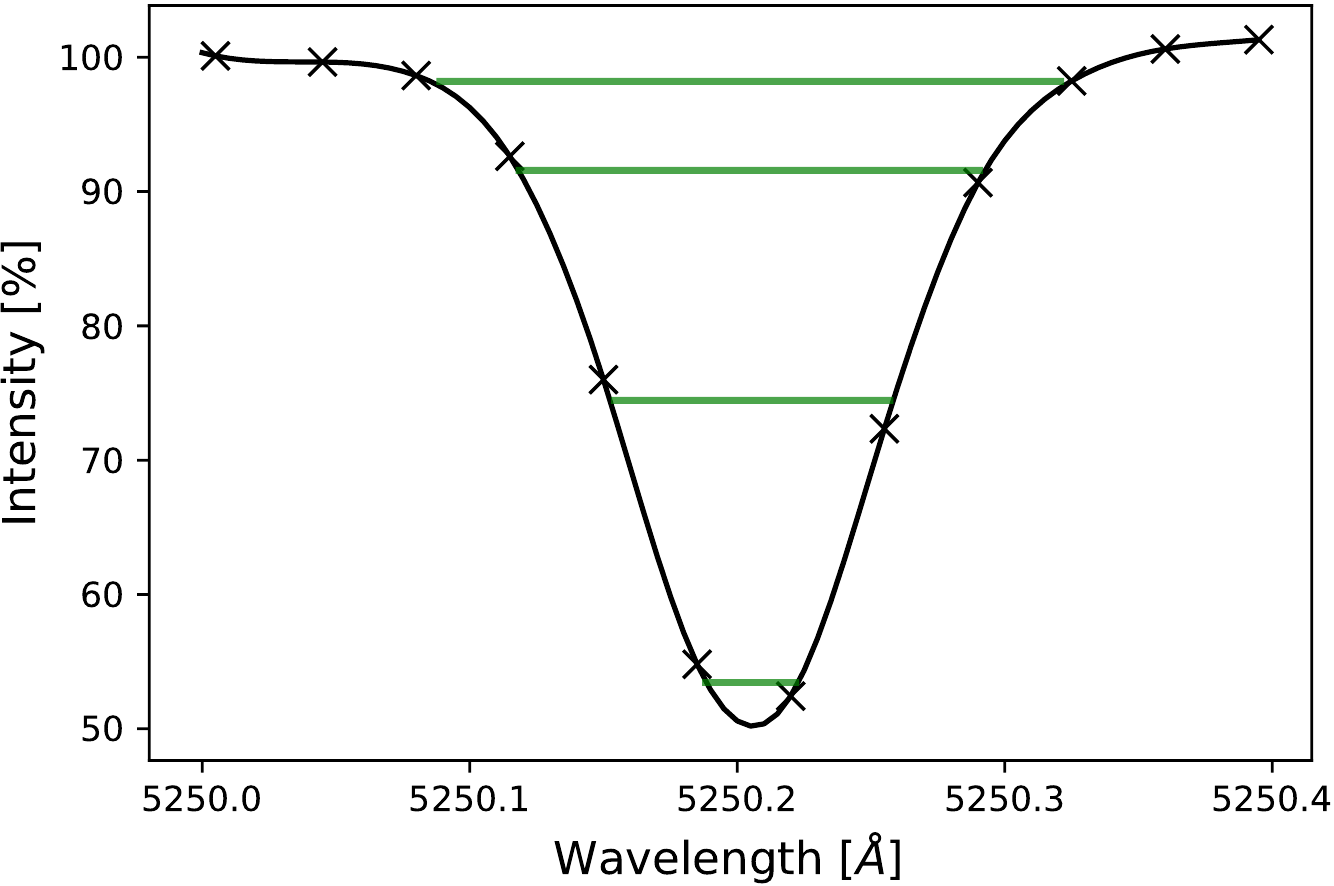}
        \caption{Line profile of the 5250.2\,\AA~line (IMaX data), where 100\% of intensity is designated as an average of the continuum. The markers denote the averaged observed line profile, and the solid line corresponds to a cubic spline interpolated profile. The horizontal green lines mark the used bisector heights that are closest to the observed wavelength points.}
        \label{fig:lineprofile5250}
\end{figure}

Doppler velocities deduced from shifts of spectral lines are well suited to obtain the average line-of-sight motion throughout the atmosphere. If we were to assume that the velocities are uniform throughout the entire atmosphere, the whole absorption line would be shifted in total. In reality, the atmospheric motions vary. The change in the velocity throughout different heights of the atmosphere leads to an asymmetric shape of an absorption line \citep{1982ARA&A..20...61D}. To illustrate these inhomogeneities, a bisector analysis can be performed. A bisector evaluates the different velocities from the different height ranges at which the spectrum is formed. For our studies we interpolated the observed data points with a cubic spline before calculating the bisector velocities. A bisector is defined as the wavelength $\lambda$ at the centre of an horizontal line, connecting the left and right wing of the spectral line at the same intensity. Figure \ref{fig:bisec} shows an averaged absorption spectrum of Fe I 6173.0\,\AA~with the bisector values. The height-dependent Doppler velocities $v_{\textrm{LOS}}$ are computed from that wavelength by
\begin{align}
        v_{\textrm{LOS}}=c\cdot \frac{(\lambda - \lambda_0)}{\lambda_0},
\end{align}
with $c$ being the speed of light. To determine the rest wavelength $\lambda_0$ , we calculated the temporal average of the line core wavelength for the whole field of view. An averaged line profile of the Fe I 5250.2\,\AA~line and the corresponding bisector heights is shown in Fig. \ref{fig:lineprofile5250}.\\
The precision with which the position of a spectral line is measured (Doppler precision) is coupled to the spectral resolution and hence to the applied spectral sampling. As a rule of thumb, the measurement precision is about 1/20th of a spectral pixel. This value was confirmed for instance by \cite{2000A&A...357..763W}. Thus, IMaX and IBIS data are well suited for measuring the temporal behaviour of granular flows, as shown in Fig.7. To remove the five-minute oscillations of the solar surface, we applied a sub-sonic filter with a cut-off velocity of 7$\,$km$\,$s$^{-1}$ \citep{1989ApJ...336..475T} to the data.

\section{Results and discussion}
\begin{figure}[h!]
        \centering
        \includegraphics[width=0.49\textwidth]{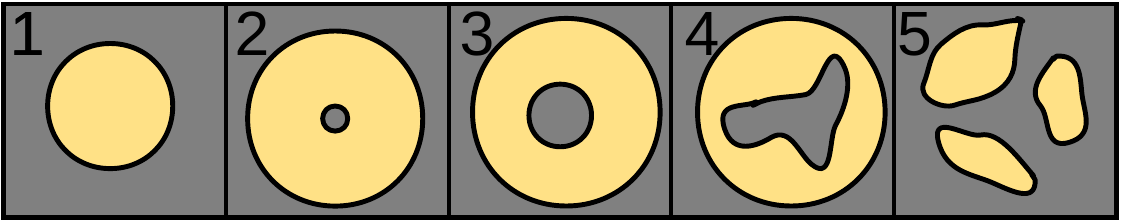}
        \caption{Visual representation of the development of an exploding granule and its disintegration by splitting into a new generation of granules. 1:~A bright granule appears and expands rapidly. 2:~Development of a decrease in intensity in the centre of the granule. 3:~Dark centre expands and downflow velocity can be measured. 4: Intergranular lane structure becomes visible. 5: Exploding granule is completely separated into a new generation of granules or fragments.}
        \label{fig:eg0}
\end{figure}

Figure \ref{fig:eg0} shows a simplified illustration of the development of a typical exploding granule until its destruction through fragmentation. We refer to these stages throughout according to the numbers given in the illustration. We performed a visual inspection of the continuum image time series of the IBIS and IMaX data and searched for exploding granules. Our criterion was the appearance of a bright doughnut-shaped structure surrounding a dark core, such as in stage 3 in Fig.\ref{fig:eg0}. 

\begin{figure}
        \includegraphics[width=0.48\textwidth]{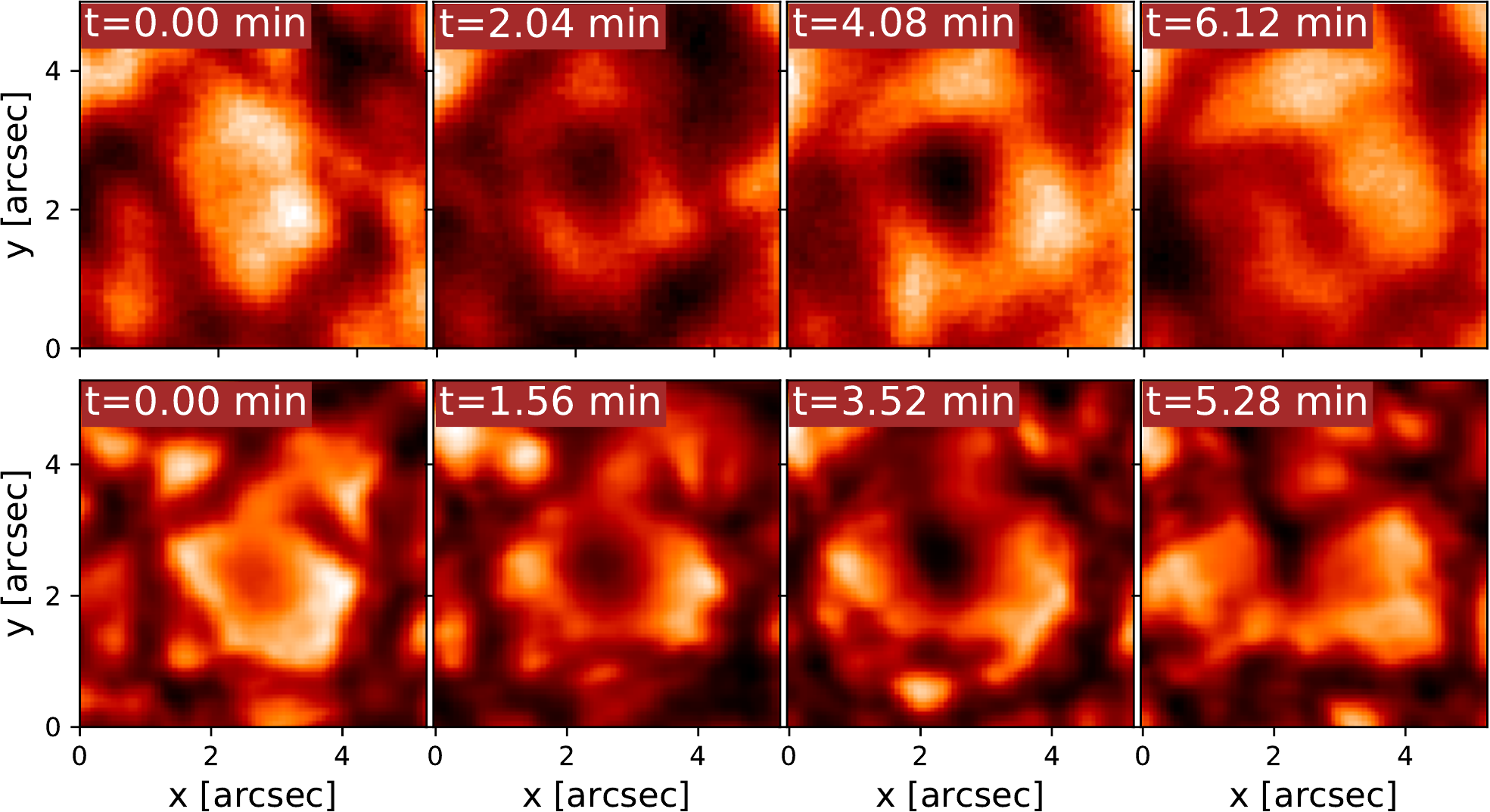}
        \caption{\textit{Upper row:} Images taken in the continuum of the IBIS data and normalised to the mean continuum intensity. Example of the evolution of exploding granule IBIS 6 (see Tab. \ref{tab:obs1}). \textit{Lower row:} Images taken in the continuum of the IMaX data and normalised to the mean continuum intensity. Example of the evolution of exploding granule IMaX 1 (see Tab. \ref{tab:obs1}).}
        \label{fig:ibis_vs_imax}
\end{figure}

The IBIS and IMaX data sets both provide a large field of view in which several exploding granule events take place within the time sequence. In Fig. \ref{fig:ibis_vs_imax} an example of an exploding granule observed with IBIS is shown in the upper row over time. The lower row shows another exploding granule, in this case, observed by the IMaX instrument. 
We selected six exploding granules from the IBIS data set and characterised them by size, expansion velocity, and lifetime. From the IMaX data, three exploding granules were examined.
In Tab. \ref{tab:obs1} all observed exploding granules are listed and named. We determined the expansion velocity, the maximum diameter an exploding granule reaches before splitting up, and the total lifetime of all IBIS events. In contrast, we only determined the maximum diameter of exploding granules observed by the IMaX instrument because the time series of IMaX has several instrumentally caused interruptions, resulting in two shorter usable sequences that therefore do not capture the events from formation to fragmentation.
According to \cite{1999ApJ...515..441H}, the lifetime of granules depends on their creation and destruction mechanisms. Their observations lead to a mean lifetime in the range between $\overline{T} = 4.49\,$minutes up to $\overline{T} = 9.23\,$ minutes. The lifetime of the observed exploding granules in this work with values between $T = 10\,$minutes and $T = 15\,$minutes is significantly higher than this mean lifetime of regular granules.

\begin{table}
        \centering
        \small
        \begin{tabular}{rrrrrr}
                \toprule
                \#& & exp. velocity& max diameter & lifetime & cadence\\
                 &&[km s$^{-1}$]&[arcsec]&[min]&[s]\\
                \midrule
                IBIS 1&&0.73 $\pm$ 0.15 & 3.1 & 12.56 &31\\ 
                IBIS 2&*&0.99 $\pm$ 0.05& 4.3 & 10.79 &49\\ 
                IBIS 3&&0.54 $\pm$ 0.13 & 3.0 & 12.45&49\\ 
                IBIS 4&&0.67 $\pm$ 0.14 & 3.1 & 14.94&49\\ 
                IBIS 5&&1.05 $\pm$ 0.05 & 2.7 & 10.99&31\\ 
                IBIS 6&*&2.13 $\pm$ 0.15 & 4.5 & 9.94&31\\ 
                IMaX 1 &*&-&4&-&29\\
                IMaX 2 &*&-&3.8 &-&29\\
                IMaX 3 &*&-&2.6 &-&29\\
                \bottomrule
        \end{tabular}
        \caption{Expansion velocity, maximum diameter, lifetime, and cadence of observation of the exploding granules from the IBIS and IMaX data. The granules marked with an asterisk form a new, independent intergranular lane with an origin in the dark centre of the exploding granule.}
        \label{tab:obs1}
\end{table}

\subsection{Newly forming downflow lanes}
New downflow lanes are typically generated by the fragmentation of pre-existing granules. During the fragmentation of the exploding granule, we typically observe dark barbs in the continuum image that develop from the dark core towards already established downflow lanes (stage 4 in Fig. \ref{fig:eg0}). These develop into new intergranular downflow lanes that eventually fragment the granule.
\begin{figure*} 
        \centering
        \includegraphics[width=0.8\textwidth]{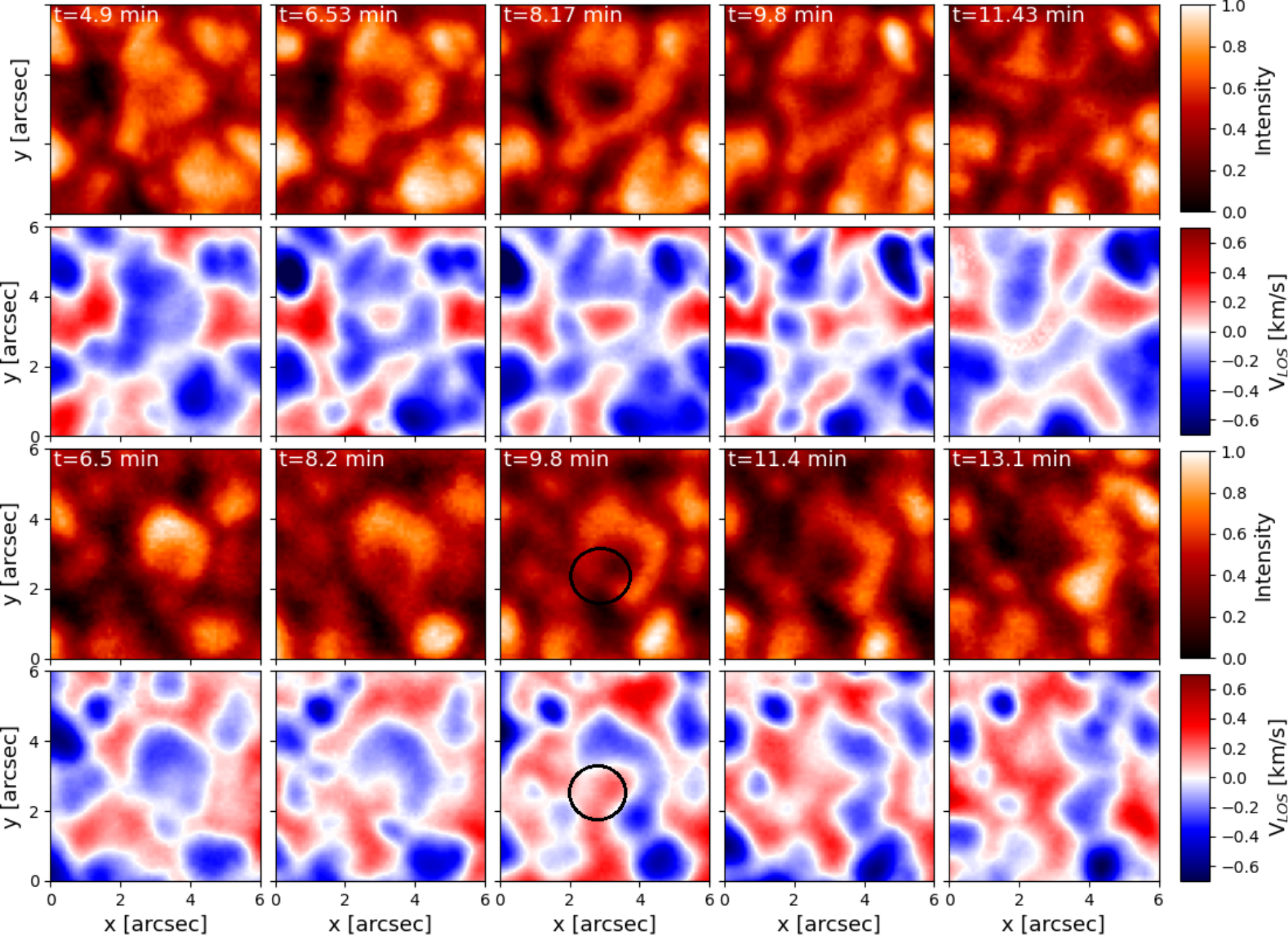}
        \caption{Evolution of two exploding granules over time. The first row shows continuum images for granule IBIS 2. Intensities are normalised to the maximum intensity in the time series. The second row shows the corresponding line-of-sight velocity maps. Positive values denote a downflow, and upflows are negative. The third row shows the evolution of exploding granule IBIS 3 in the continuum images. The last row shows the corresponding line-of-sight velocity map for this event. The black circle refers to the area mentioned in the text.}
        \label{fig:evol}
\end{figure*}

The evolution of two different exploding granules from the IBIS data set is shown in Fig. \ref{fig:evol}. The continuum and line-of-sight velocity maps illustrate the onset of the dark core and the dissolution process over several minutes.
At first inspection, granule IBIS 2 (first two rows) clearly shows a development according to the stages in Fig. 3, whilst the dark core in granule IBIS 3 (last two rows) develops very close to an established intergranular lane. A closer look at the examples shown in Fig. \ref{fig:evol} reveals a difference in the formation of new downflow lanes. Granule IBIS 2 shows a doughnut-shaped exploding granule in the continuum light and also in the velocity map. The dark core of this granule evolves in the centre of this. In contrast, the downflow in granule IBIS 3 appears to develop starting from the already established intergranular lane at the edge of the granule, and it later visually forms a bridge to the central dark core (see the area marked with a black circle). At $t=3.27\,\textrm{min,}$ this granule appears to be already split up in the velocity maps.

Comparing the continuum images and the line-of-sight velocities for all granules, we find that for some granules, the intensity drop in the centre of the granule precedes the downflow in the velocity map while it coincindes for others, and we also observe cases in which the downflow occurs first.
The dissolution shows a striking difference: Whereas the larger granule IBIS 2 fragments by being split up by a newly forming intergranular lane, granule IBIS 3 seems to shrink, with the dark core fading into the close pre-existing intergranular lane. When the entire data set is classified, we find that all of the exploding granules with a diameter larger than 3.8$\,\arcsec$ form an independent new intergranular lane that develops from the dark core. Exploding granule IMaX 3 also shows this development into an independent intergranular lane, but only reaches a maximum diameter of 2.6$\,\arcsec$. All other smaller granules disappear by fading and disappearing into the surrounding intergranular lanes.

Granules IBIS 2 and IBIS 3 show a maximum downflow of 0.27 km$\,$s$^{-1}$ and 0.22 km$\,$s$^{-1}$, respectively, within the dark core. For the average of the maximum line-of-sight velocity in the dark core for the six observed granules in the IBIS data we obtain a mean downflow velocity of 0.24$\,\pm$ 0.05 km$\,$s$^{-1}$.

We also analysed whether the newly formed downflow lane at the dark core showed a similar behaviour as regular intergranular lanes. We therefore studied the height-dependent variation of bisector velocities.
\begin{figure*}
        \centering
        \includegraphics[width=0.8\textwidth]{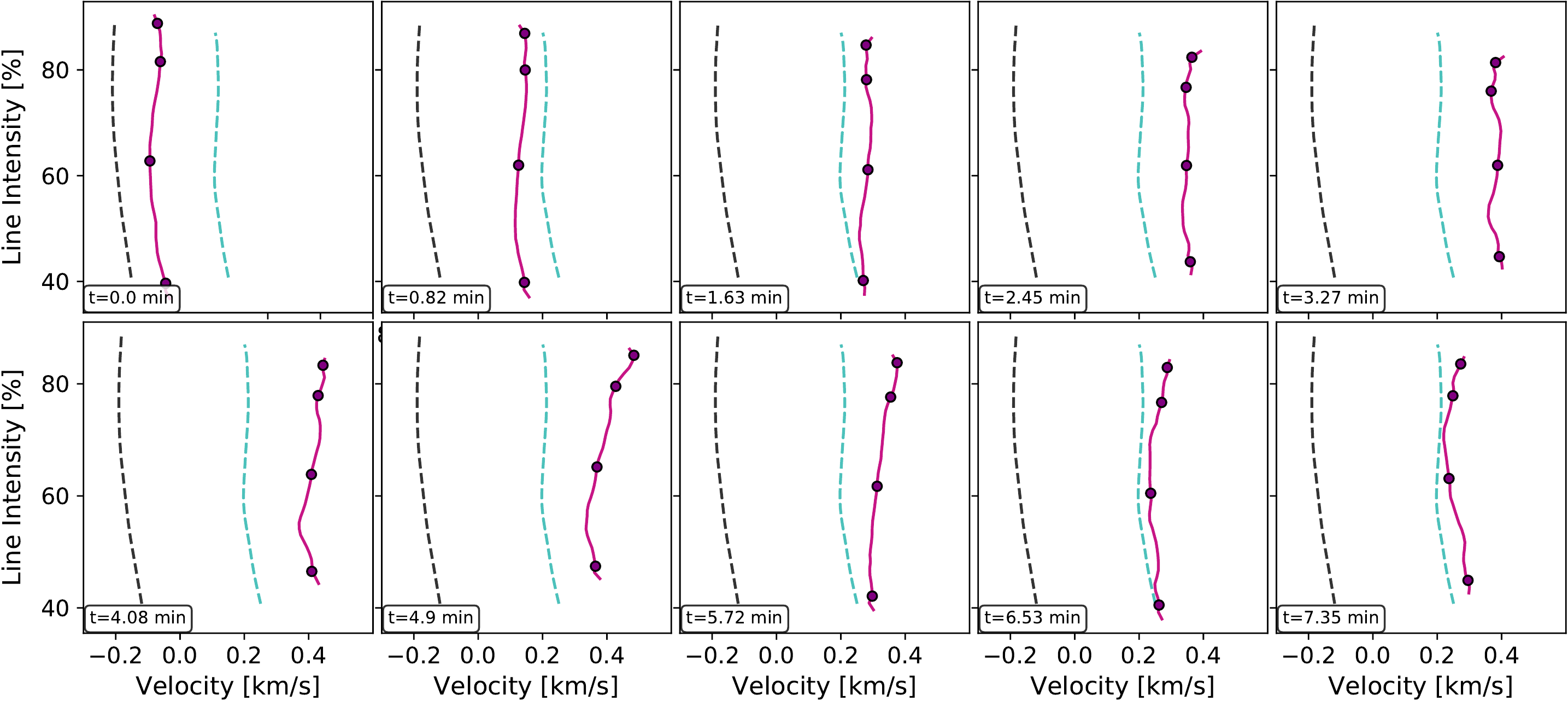}
        \caption{Time evolution of a bisector (pink solid line) measured in the dark core of exploding granule IBIS 2. Black dotes correspond to the bisectors measured at the spectral locations closest to the observations. The dashed black lines denote the mean bisector of regular granules, and the dashed turquoise line the mean bisector of the regular intergranules. Time evolves from left to right and continues in the second row.}
        \label{fig:GRIG}
\end{figure*}
To this end, we first distinguished granules and intergranular lanes by using intensity maps over the field of view in the continuum and over time. With a limit of $\pm$5$\,\%$ of the mean intensity for granular and intergranular areas, respectively, we determined averaged bisector shapes for both cases. In Fig. \ref{fig:GRIG} these averaged bisectors are represented by the dashed lines. The dark core and subsequent newly forming downflow lane within the exploding granule is represented by an average of 9 pixels from the centre of granule IBIS 2. From time step t=0$\,$min to t=2.45$\,$min, in the phase in which the dark core starts to form (stage 2 in Fig. \ref{fig:eg0}), we observe that the velocities through the photosphere transform with time relatively uniformly from an upflow to a downflow. The downflow velocities in the dark centre of the exploding granule are somewhat higher at all heights than the averaged velocity of the typical intergranular lanes. When the dark core becomes visible in the continuum images (stage 3 in Fig. \ref{fig:eg0}), the bisector shows a slight bend towards higher downflow velocities in the lower atmosphere. At t=6.53$\,$min, the bisector comes close to the shape and velocity values of typical intergranular downflow lanes. At the same time, the continuum images show the new fully developed intergranular lane (stage 5 in Fig. \ref{fig:eg0}).

Images at the wavelength at which the line intensity is higher than about 70\,\% show a clear granulation pattern with bright granules and dark intergranular lanes and correspond to the lower photospheric atmosphere. The bisector velocity values obtained at this intensity level accordingly show a downflow in areas of intergranular lanes, whereas bisector values analysed closer to the line core carry information of the higher photospheric atmosphere. From an intensity level of about 70\,\%, a reversed granulation signature appears in which upflows replace the downflows in the intergranular structure.

\subsection{Height dependence} \label{Height_dependence}
\begin{figure}[h!]
        \begin{minipage}{.5\textwidth}
                \includegraphics[width=\textwidth]{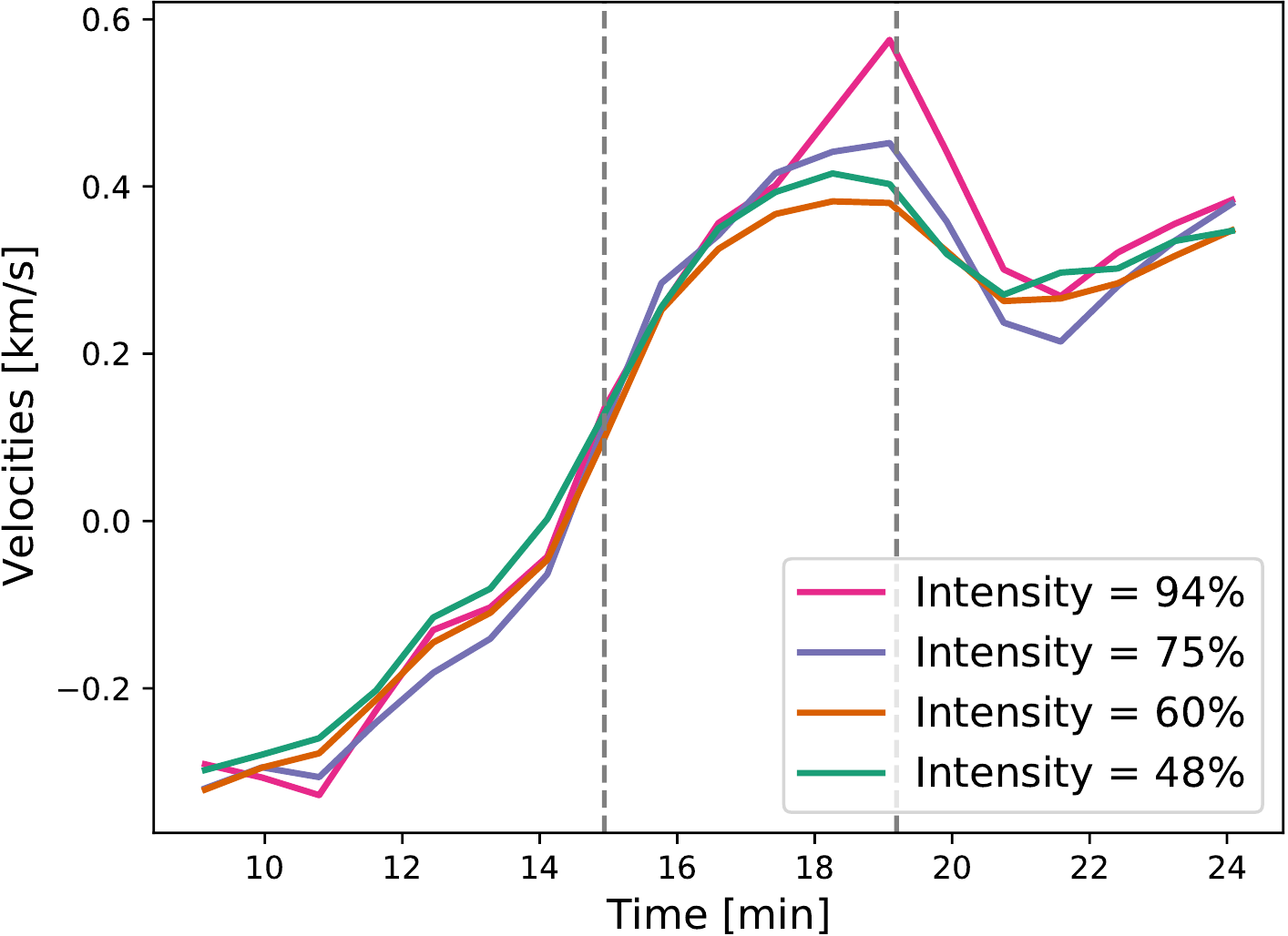}
        \end{minipage}
        \begin{minipage}{.5\textwidth} 
                \includegraphics[width=\textwidth]{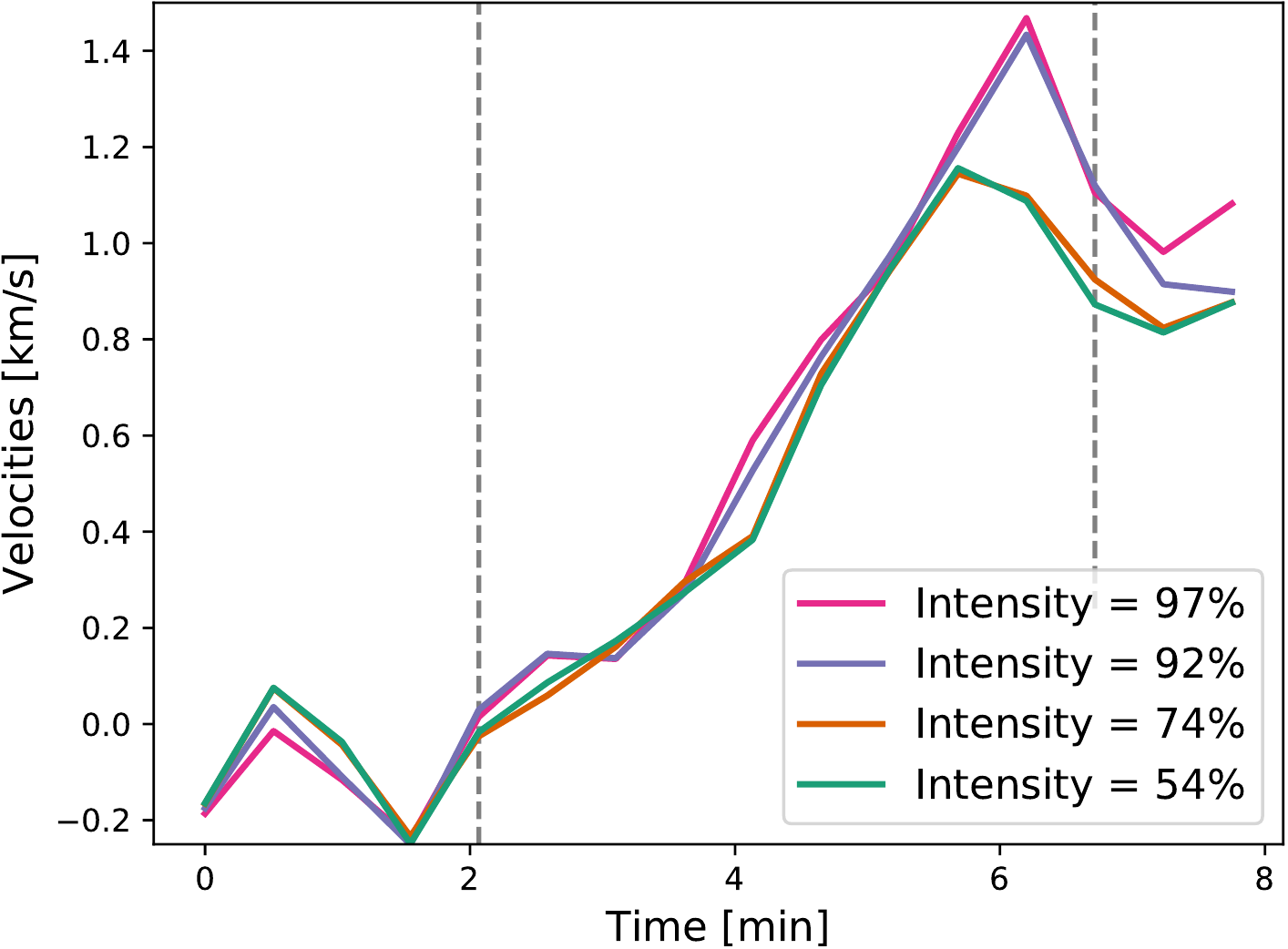}
        \end{minipage}
        \caption{Velocities at different intensity levels in the spectral line at which the bisectors are calculated  (corresponding to different atmospheric heights) for the centre of the dark core plotted over time. The upper plot shows the velocities of IBIS 2 granule, inferred from the 6173.0\,\AA~line. The lower plot shows the velocities of IMAX 3 granule, inferred from the 5250.2\,\AA~line. The left dashed line marks the time at which the dark core is visible for the first time, and the right dashed line marks the time at which the exploding granule is split into a new generation of granules.}
        \label{fig:vel_gr}
\end{figure}

When we assume that a dark core forms as a result of buoyancy braking and radiatively cooled gas returning to the solar surface from higher in the atmosphere, we would expect a delay in time for downflowing material when the onset of the downflow at different heights in the atmosphere is timed. This has been observed by \cite{2017ApJ...836...40O}, who found a delay of less than 30\,s for a height difference of about 120\,km by using slit spectropolarimetric data. Because our data have a lower cadence, this delay cannot be resolved adequately in the IBIS or IMaX data. However, we observe another phenomenon concerning the time evolution of the height-dependent downflow velocity. The time at which the maximum downflow velocity of the dark core occurs at any given height seems to possess a shift in time for some of the studied exploding granules. The lower in the atmosphere of the Sun, the later the development of the maximum of the downflow inside the dark core. This delay was only detected for the two largest granules in our IBIS sample, which are also the only two IBIS granules that form a new intergranular lane by fragmentation.
The four smaller exploding granules from IBIS did not show this effect. Instead, the time at which the velocities reach their maximum at the different heights is nearly identical. In Fig. \ref{fig:vel_gr} the temporal evolution of the velocities of the dark cores of two granules are shown. We show the velocities of different intensity levels in the spectral line at which the bisectors are calculated (corresponding to different heights). The temporal shift of the maximum velocity through the solar atmosphere of the granules IBIS 2 and IMaX 3 is shown. The maximum downflow velocity of IBIS 2 at line intensities of 48~\% and 60~\% (high photosphere) is reached earlier than at 75~\% and 94~\% (lower photosphere), with a delay of about 48\,s. This delay was also found with about 61\,s for the IBIS 6 granule (see the appendix, also for further examined granules). IMaX 3 is a smaller granule with a maximum diameter of 2.6$\arcsec$. However, in contrast to the small IBIS granules, IMaX 3 granule forms an intergranular lane from its central dark core. Fig. \ref{fig:vel_gr} shows that this granule also exhibits the described time shift in the maximum downflow velocity. The time shifts $\Delta$t$_{v_{max}}$ measured by the Fe I 5250.2\,\AA~lines of IMaX are about 20\,s for IMaX 3 and 23\,s for IMaX 2. For the granule IMaX 1 we cannot make a comparable statement for the velocity development because this granule did not show a well-developed velocity maximum, see \ref{fig:bisec_imax}. However, for every other granule forming a new intergranular lane from their dark core, we find a time shift.

The differences between the IBIS results might arise because the data containing exploding granule IBIS 6 were much noisier than the data for granule IBIS 2. In addition, the data sets have a different cadence (see Table \ref{tab:values}).\\

A comparison of the IBIS and IMaX data might also not be straightforward. Both lines are photospheric, but they are still not directly comparable. For example, the line core of Fe I 5250.2\,\AA~is approximately formed at a height of 440 km and the Fe I 6173.0\,\AA~core forms at a height of 387 km over $\tau_{500}$=1, as confirmed using local thermodynamic equilibrium (LTE) simulations of \cite{1998A&A...329..721S}. Additionally, both curves originate from different spectral resolution observations. \cite{2019A&A...622A..34S} showed that an increased spectral resolution results in a more strongly developed C-shape of the bisector. Because the IBIS data have a higher spectral resolution than the IMaX data, the time shift of the maximum velocities may also be less pronounced in the IMaX observations.

\subsection{3D MHD simulations}\label{simulations}

\begin{figure*}
	\centering
	\includegraphics[width=\textwidth]{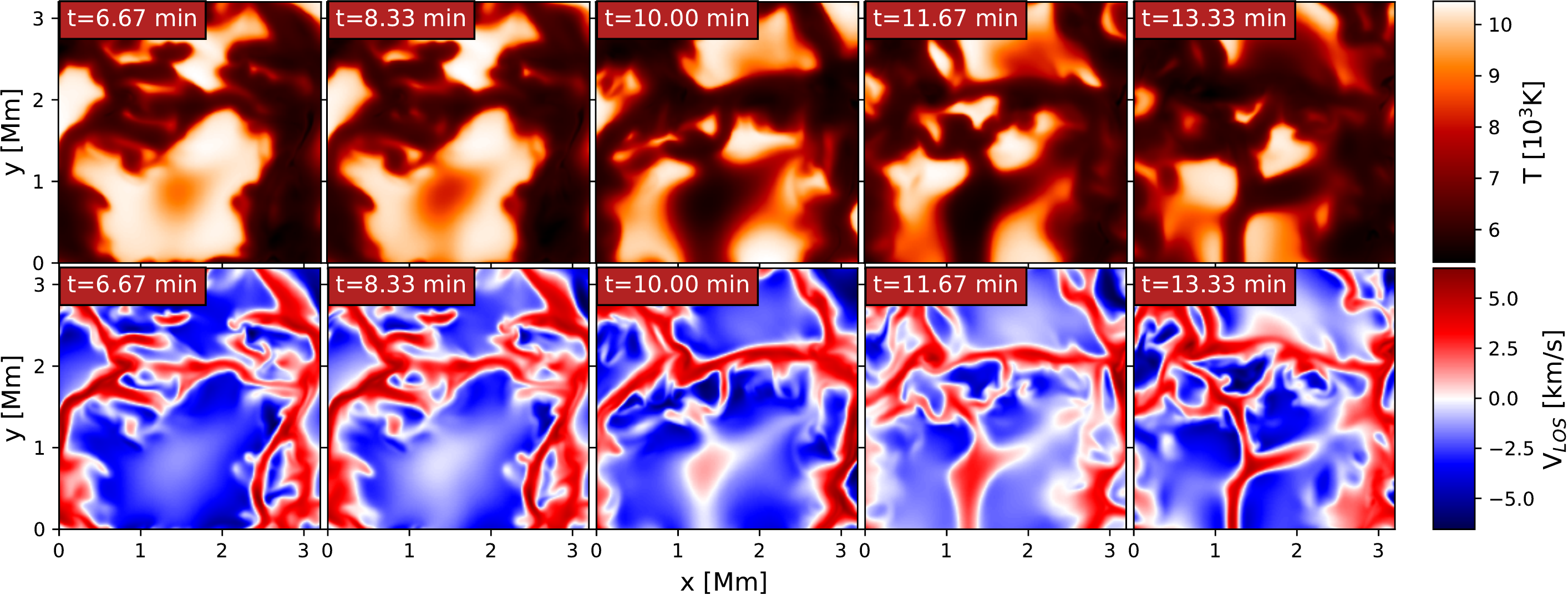}
	\caption{Time evolution of an exploding granule obtained with simulations of the \textsc{M\tiny{ANCHA}}3D code. The first row shows the temperature at height 0\,km, and the second row shows the corresponding vertical component of the velocity. The times mentioned here are the relative times of the simulation with t=0\,min as the starting point of our extracted time series.}
	\label{fig:img_MANCHA}
\end{figure*}

\begin{figure}
        \includegraphics[width=0.46\textwidth]{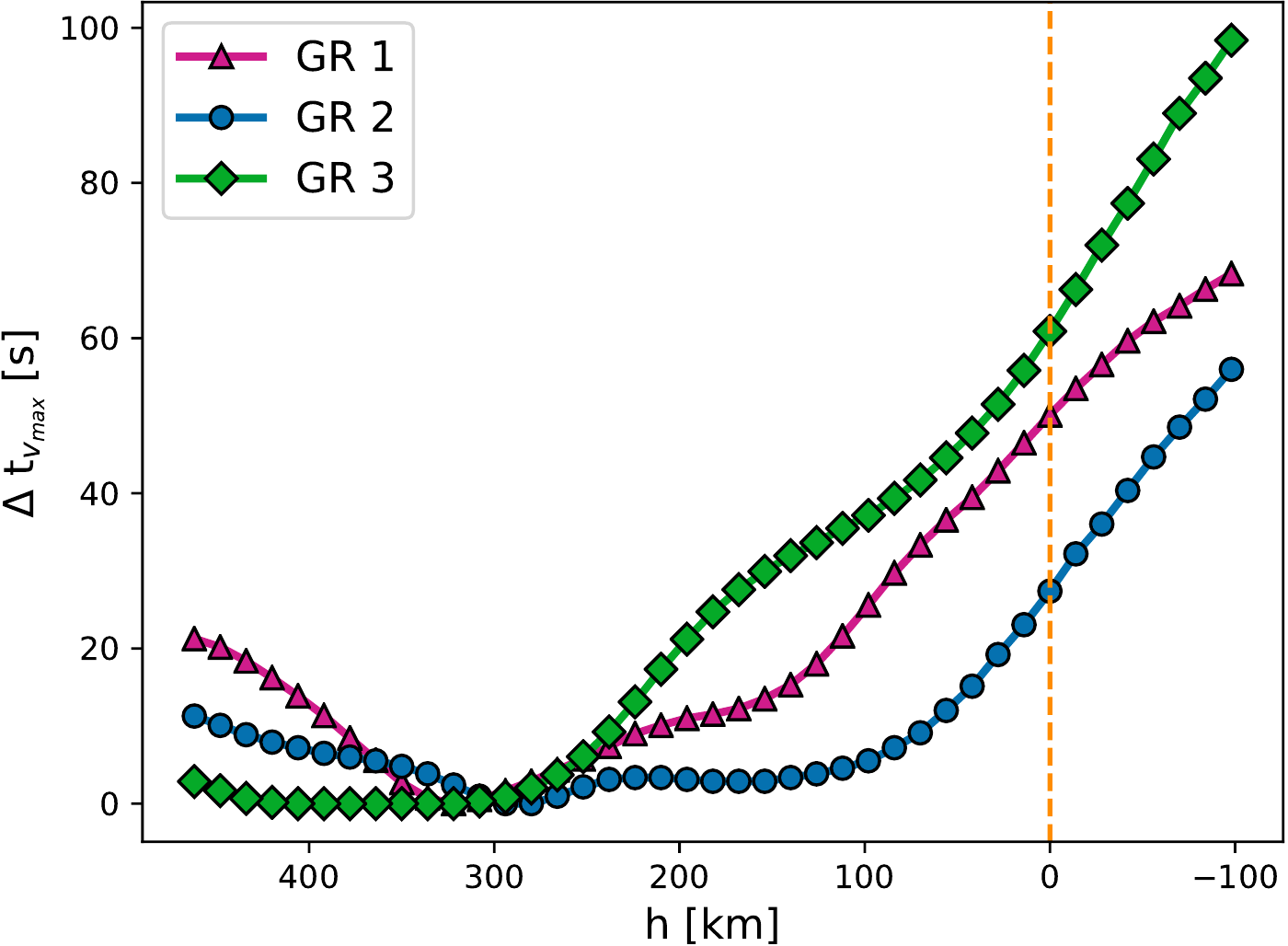}
        \caption{Time delay $\Delta$t$_{v_{max}}$ plotted over the height in the simulated solar atmosphere from 100 km below the photosphere to about 476 km above for different simulated granules. The dashed orange line indicates h=0~km.}
        \label{fig:diffGr}
\end{figure}

Three-dimensional radiative-magnetohydrodynamic (RMHD) simulations provide deeper insight into the physical processes in the solar atmosphere through a comparison of observations and comprehensive theory. For this purpose, we additionally analysed \textsc{M\tiny{ANCHA}}3D simulations \citep{2017A&A...604A..66K}. We used these simulated data to verify whether the velocity shift mentioned in Sec. \ref{Height_dependence} also occurs in atmospheric models. In the simulations the time-dependent, non-ideal, and non-linear equations of the single-fluid RMHD theory are solved on a 3D Cartesian grid. It is assumed that  LTE holds. The effects of the ambipolar diffusion due to the partial ionization of the plasma are included. The computational domain covers a height range from 0.95\,Mm below the surface up to the lower chromosphere at approximately 1.4\,Mm above the solar surface.  The simulation contains a ($5.8 \times 5.8 \times 2.35$)\,Mm$^3$ box with the grid points distributed uniformly  in the horizontal directions with 20\,km and in the vertical axis with 14\,km. Details about the simulation setup can be found in \cite{2018A&A...618A..87K}. We extracted time sequences at a cadence of 20\,s from the simulations.

In the simulated data set we found three striking examples of large exploding granules. In Fig. \ref{fig:img_MANCHA} the evolution of one of these granules is shown in temperature at a height of 0\,km (geometrical height corresponding to the mean optical depth unity) in the photosphere, together with the vertical component of the velocity at the same geometrical height. The exploding granule in the simulation exhibits a temporal behaviour that is similar to that of the observed exploding granules that develop a new intergranular lane.

We are again interested in the velocity evolution within the dark core at the centre of the exploding granule. We calculated for each height the time of the maximum downflow during the formation of the intergranular lane and determined the temporal difference $\Delta t_{v_{max}}$ between these respective highest velocities (see the discussion of this time shift in Sec.~\ref{Height_dependence}). In Fig.~\ref{fig:diffGr} this time delay of the maximum downflow velocity ($v_{z}$) versus the height within the simulation box for the three examined exploding granules is shown. The data point at $\Delta t_{v_{max}} = 0\min$ corresponds to the first maximum that occurs in time . In all three cases, it is at a height of approximately 300~km. In case of GR 3, the maximum is reached at the same time for heights between about 300 and 400~km. Although the shape of the lines varies, they all follow the same trend of an increasing delay of the maximum downflow when progressing into the lower atmosphere. The time delay is not linear and proceeds well below the solar surface. 
We do not intend a comprehensive 1-to-1 comparison to the observations, but we endeavour to estimate the imprint that this velocity evolution would have on the observed spectral lines. We also studied the effect of the resolution on the appearance of the exploding granule. We used the MANCHA-RAY code (Vitas \& Khomenko, 2021, in preparation) to compute synthetic spectra for the Fe I 6173\,\AA~line, which is the same line that was observed by the IBIS instrument, from the simulations. The spectra were spatially and spectrally degradedd. For the spatial blurring we use oint spread functions (PDFs) approximated by Gaussians with a differing full width at half maximum (FWHM) of 120\,km, 180\,km, and 240\,km. We binned the spectral profiles to match different pixel sizes of a camera (e.g. a binning to $0.098$\,arcsec corresponds to the IBIS camera pixel size). The spectra were blurred with a Gaussian of FWHM=25\,m\AA, which is a good approximation of the spectral resolution of the IBIS instrument (see~\cite{2008A&A...481..897R}), and we then sampled the profile at a reduced spectral resolution of 30\,m\AA, which is similar to the spectral resolution in our IBIS data.

\begin{figure*}
    \centering
    \includegraphics[width=0.98\textwidth]{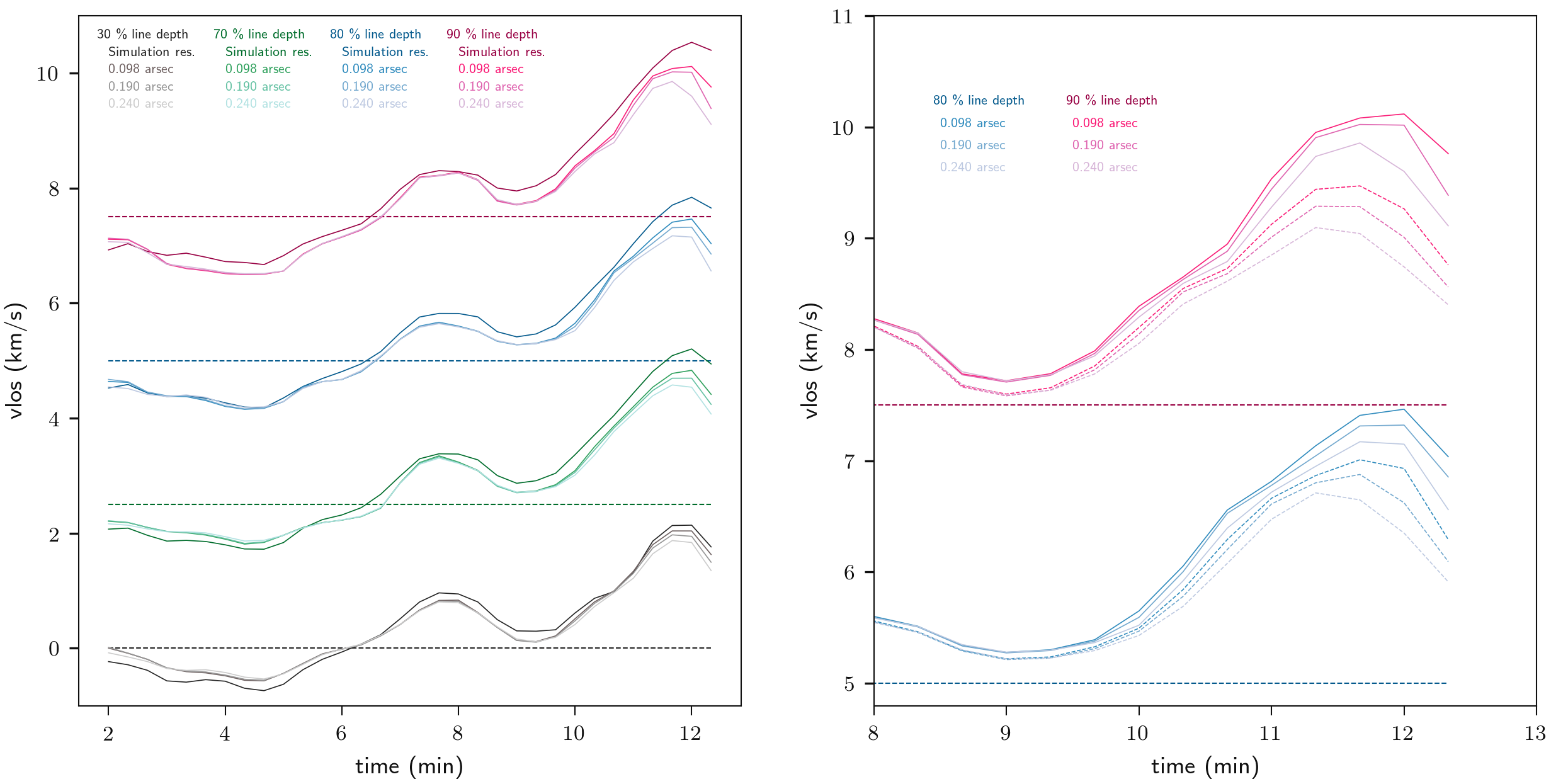}
    \caption{Line-of-sight velocities obtained with a bisector calculation for varying intensity levels of spectral lines synthesised from the \textsc{M\tiny{ANCHA}} atmospheres. Left: Velocities for non-degraded spectra synthesised at the simulation resolution. The degraded spectrum has undergone spatial and spectral blurring, was re-sampled to an instrumental spectral resolution, and was re-binned to different pixel sizes (see text for details). The curves for the different percentages of the line depth are offset by 2.5\,km/s for clarity. Dashed lines correspond to 0\,km/s for the different percentages of the line depth. Right: Bisector velocities calculated from synthesised and degraded spectral profiles as in the left image, now shown for only two intensity levels with a spatial blurring employing a Gaussian of FWHM of 120\,km (solid lines, same as in left panel) and an FWHM of 240\,km (dashed lines). The curves for the different percentages of line depth are offset by 2.5\,km/s for clarity. The horizontal dashed lines correspond to 0\,km/s for the different percentages of the line depth. }
    \label{fig:tdevs}
\end{figure*}
   
\begin{figure*}
    \centering
    \includegraphics[width=0.98\textwidth]{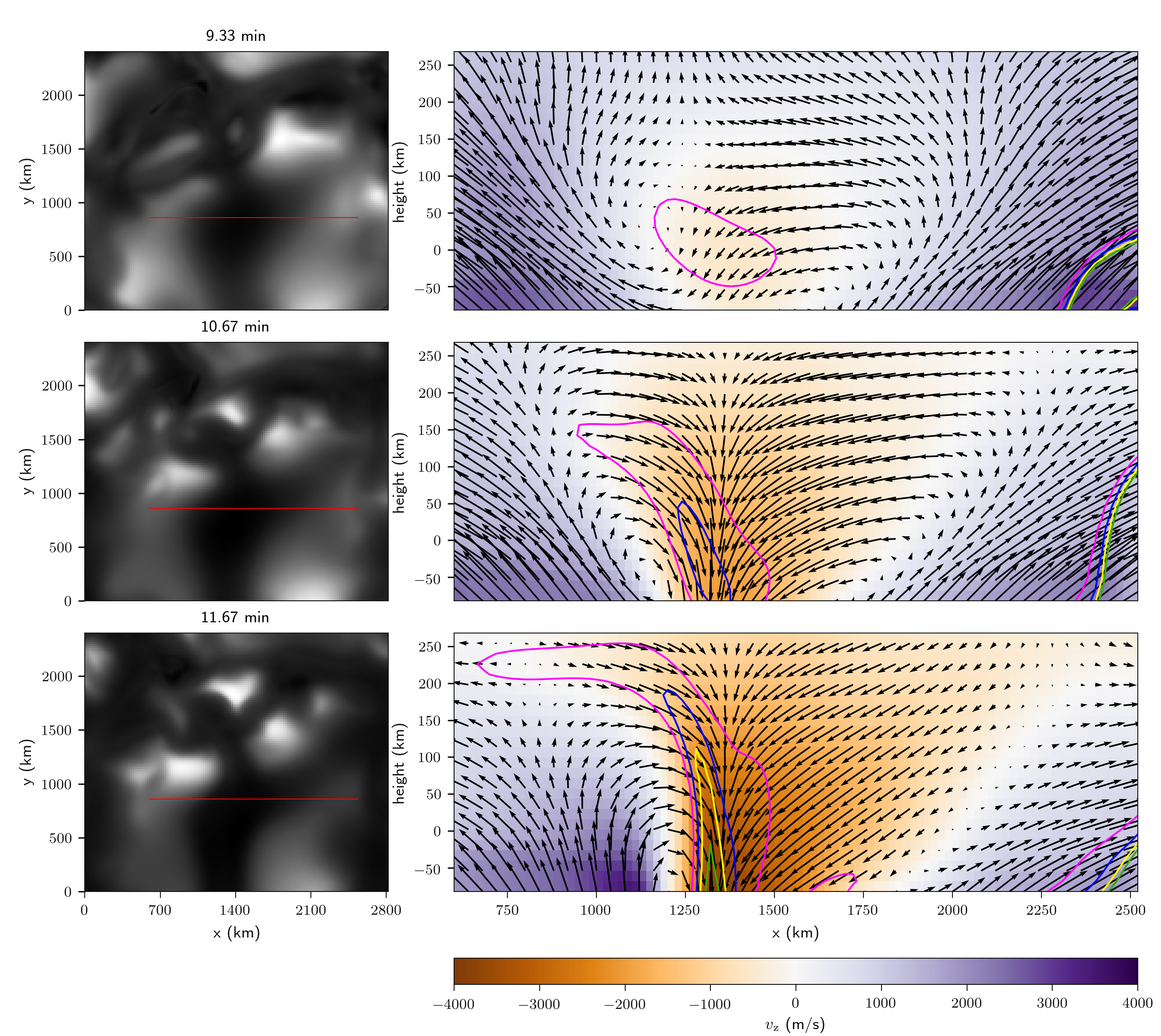}
    \caption{Left column: Temperature maps at 0\,km height of the simulation box showing the development of the dark core within the exploding granule shown in Fig.~\ref{fig:img_MANCHA} for different times. The same time t=0 is chosen for both plots. Right column: Cut through the $x-z-plane$ of the atmosphere along the red line in the images on the left. Coloured background shows the v$_z$ velocity, and the black arrows represent the projection of the velocity vector onto the  $x-z-plane$. Additional contours show the magnetic field strength at levels of 10\,G (pink), 20\,G (blue), 30\,G (yellow), and 40\,G (green).}
    \label{fig:funnel}
\end{figure*}

We produced similar plots as for the observations by calculating the bisector velocities for different intensity levels in the atmosphere (corresponding to different heights in the atmosphere).
In Fig.~\ref{fig:tdevs} we show the line-of-sight velocities obtained through a bisector calculation at the location of the dark core for GR 3. The left image shows the bisector velocities calculated from spectral profiles synthesised from the simulation as well as from spectrally degraded (with a Gaussian of FWHM=25\,m\AA) and spatially degraded (with a Gaussian of FWHM=120\,km) synthesised spectra for different pixel sampling. 
The velocities calculated at the simulation resolution show a slight shift in the time at which the maximum velocity is reached at different intensity levels. The deeper in the atmosphere, the later the maximum. This is in line with what was found in the velocity evolution shown in Fig.~\ref{fig:diffGr}, but the delay is much shorter (now only about 8\,s) and therefore less noticeable. This is most likely due to observational degrading and also because the Fe I 6173~\AA~lines with their formation region only cover a specific range of heights in the simulated atmosphere (cf. Sec. \ref{Height_dependence}).

The velocities calculated from non-degraded spectra are generally higher  than for the degraded spectra. In velocities from degraded spectra,  a variation depending on pixel sampling size in the curves becomes visible only after 12 minutes. Here, the maximum velocity inside the dark core is reached and shows a higher value for higher resolution (smaller pixel size), especially in the lower atmosphere. 
At an intensity level of approximately $30\%$ of the line depth (green curves), the curves show less variation in the velocity profile when the pixel size is increased compared to what is seen at a line depth of $90\%$ (lower in the atmosphere).
 
The stagnation seen at about 8 minutes in Fig.~\ref{fig:tdevs} (left panel), which leads to the second minimum during the establishment of the downflow, is seen at all heights and spatial resolutions. This means that it presumably affects the entire structure along the line of sight. This might be caused by an initial collision of the downfalling material with the denser upflowing material within the newly forming downflow lane. After the stagnation phase, a downflow can finally establish itself. 
The overall shape of the velocity curve found from the degraded spectra is similar to the curves shown in  Fig.~\ref{fig:vel_gr}, for example, including a stagnation point or hump seen at 14\,min (upper panel) and 3\,min (lower panel). Most observed granules also show the development of the highest downflow velocity at a delay in the deeper layer, similar to what is found from the synthetic spectra. An exception, as previously mentioned, is the exploding granule IMaX 1 (see Fig.~\ref{fig:bisec_imax}). The amount of this delay might show a large distribution in exploding granules depending on factors such as amplitudes of the velocities within the newly forming downflow lanes, the size of the exploding granule, and the lifetime. Given the few observational and simulation examples of exploding granules, we did not attempt such a statistic.
 
When we compare the calculated bisector velocities derived from synthesised spectra spatially degraded with Gaussians of two different FWHMs (see Fig.~\ref{fig:tdevs}, right panel), the increase in spatial blurring leads to a decrease in velocity magnitude and shifts the time at which the maximum downflow is achieved to an earlier time. The increased blurring therefore has a similar effect as the increase in binning, as expected from this further spatial degradation of the spectra.
 
The findings gathered from Fig.~\ref{fig:diffGr} and Fig.~\ref{fig:tdevs} can be explained by the different pixel sizes at which the fine structure and temporal evolution of the newly forming downflow lane are sampled. Fig~\ref{fig:funnel} shows  a sideview ($v_x - v_z$ plane) of the newly forming downflow lane at three time steps. At 9.33\,min into the sequence, the stronger downflow is initialised in a thin region (100\,km width above and below the simulated solar surface). The downflow then rapidly expands into the upper and lower regions and increases its speed. This is counterintuitive to the idea that the downflow first occurs in the upper granular convection region, but explains the delay noted in Figure~\ref{fig:tdevs} for the first minimum.  At 11.67\,min, the downflow is established throughout the lower photosphere and below the solar surface and appears as a funnel structure with higher speeds at the centre of the funnel. The edges of the funnel show turbulent flow fields, as seen
in the overplotted velocity arrows. Vortex flows appear in the regions that separate upflows and downflows. The contours of the magnetic field strength show a magnetic field of 10\,G that forms at the location of the developing downflow (upper panel in Fig~\ref{fig:funnel}). For this cut, it is roughly co-located with the area exhibiting the initial downflow. As the downflow becomes more pronounced and affects a larger area within the granule (middle and lower panel of Fig. \ref{fig:funnel}), so does the magnetic field. It increases within a larger area, still co-spatial to the downflow region, and reaches about 40\,G. This behaviour is similar to the findings of~\cite{2018ApJ...859..161R}, who described an intensification of the magnetic field up to 800\,G within the newly forming downflow lane of exploding granules. We find a much lower magnetic field strength, however. As the magnetic field values are very low, similar fields would be difficult to observe and are certainly beyond the capability of the observations we discuss here. The two other exploding granules found in the simulations also show a weak field ($<$100\,G) within the new downflow lane, but it is now initially detected above the new downflow region before it expands within the funnel.\\
Describing the newly forming downflow lane as a funnel can then also explain the differing behaviour with height of the curves in Fig.~\ref{fig:tdevs}. In the higher atmosphere, where the funnel is spatially broad and wide, it shows a homogeneous atmosphere to a larger extent. The spatial resolution becomes less relevant as the same structure or atmosphere is observed within the funnel structure. 
However, the lower in the solar atmosphere, the narrower the funnel structure. It also develops a fine structure in the strength of the downflow: the downflow in the core has accelerated more strongly. The larger pixel sampling size mixes the reduced downflows and even upflows at the edges of the funnel with the strong downflows that occupy a narrow channel. With better spatial resolution, the atmospheric component with stronger downflow starts to dominate the sampled atmosphere. This explains the higher downflow velocities that found with high resolution.\\ 
The signature of the temporal delay of the maximum downflow is therefore a sign of the combined effects of spatial fine structure and velocity gradients along the line of sight within the funnel structure. This in turn also means that the bisector analysis of the velocity at the dark core of exploding granules can serve as an indicator of the spatial and line-of-sight inhomogenities within the newly forming downflow lanes.

\begin{figure*}
    \centering
    \includegraphics[width=0.95\textwidth]{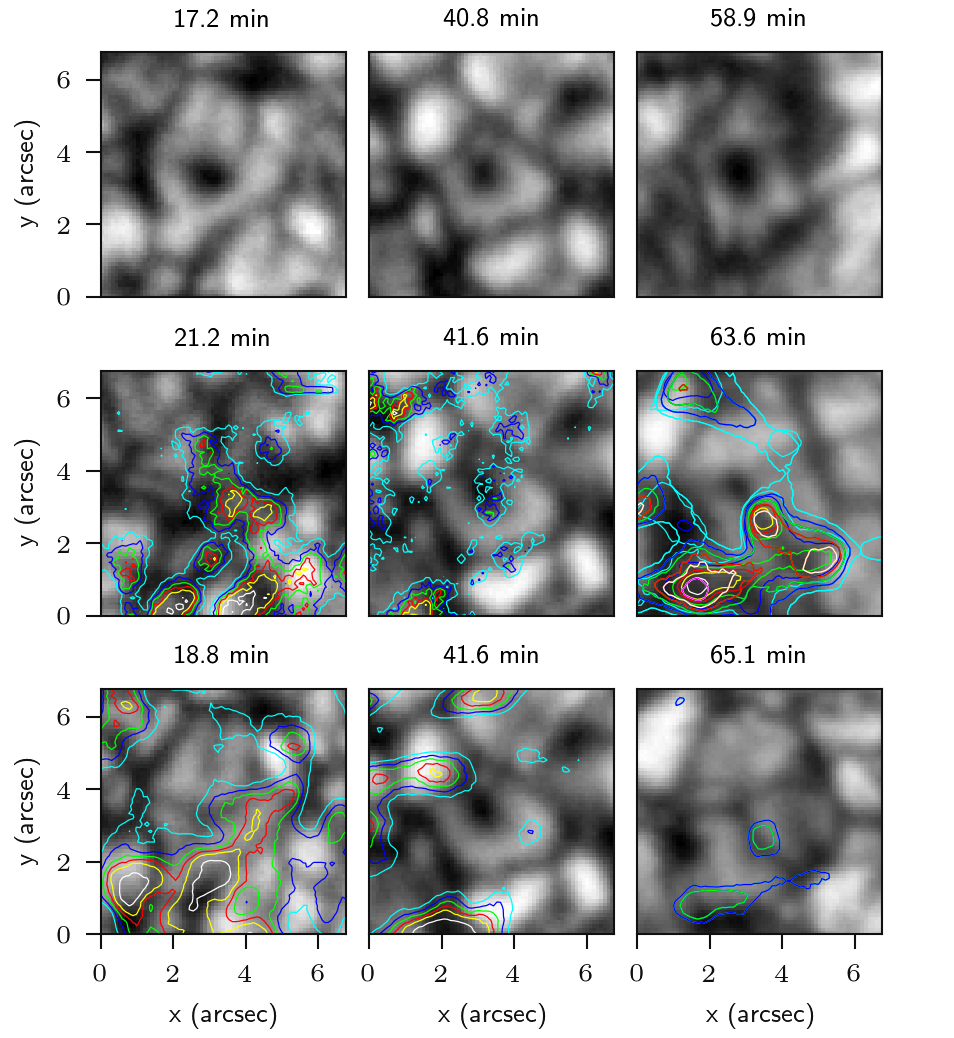}
    \caption{Top row: Continuum images of three different exploding granules during the formation of the dark core recorded with the IBIS instrument. Second row: Image in the background again shows the continuum image. Areas with a wavelet power for normalised line-core intensity oscillations in the Ca II $8542$\,\AA~(first two exploding granules) and Na I $5896$\,\AA~(last granule) lines are marked by contours, with the averaged power in the period range of 120\,s to 190\,s. Colours correspond to different percentages of the maximum power in the field of view (cyan 20$\%$, blue 30$\%$, green 40$\%$, red 50$\%$, yellow 60$\%$, white 80$\%$, and fuchsia 100$\%$). Lower row: Line-of-sight velocity oscillation power in the same format as the second row.}
    \label{fig:wavelet}
\end{figure*}

\subsection{Intensity and velocity oscillations}
In the previous section, the dark core and newly forming downflow lane were shown to be locations in which the atmosphere exhibits a varying velocity structure that is more turbulent in some areas. It is plausible that these locations are then susceptible to local transient pressure variations, hence implying that the newly forming lanes are possible sources for acoustic events.\\ In addition to the photospheric Fe I lines, IBIS acquired data in the high photospheric line of Na I $5896\,$\AA~on 14 October and of the chromospheric Ca II $8542\,$\AA~on 12 October (see Tab.~\ref{tab:obs1}). We applied a wavelet analysis~\citep{1998BAMS...79...61T} on the spectral core intensity and on the line-of-sight velocity fluctuations of these spectral lines. We searched for enhanced wavelet power in the region of the exploding granule during the onset of the dark core up to the fragmentation.\\
In Fig.~\ref{fig:wavelet} we show three of the exploding granules observed with the IBIS instrument, one granule per column. The first image in the second row shows the obtained wavelet power averaged over the 120\,s to 190\,s range for the intensity fluctuations in the Ca II $8542\,$\AA~line. The largest power is obtained in the intergranular lane at the lower right boundary of the exploding granule. At $\approx$\,4 minutes after the dark core was clearly seen (see the first panel in the upper row), the wavelet power within the exploding granule area reaches a large extent and almost covers the entire exploding granule. At this time, the granule is in the process of splitting up, and the obtained power is almost comparable to what is detected in the intergranular lanes. The second panel in the second row is obtained in the same manner, but for the exploding granule shown in the second panel in the upper row. Again, most of the power is concentrated in the intergranular lane. We observe a patch of enhanced power in the region of the dark core before the onset of the actual splitting of the granule. This is interesting because this is thought to be a particularly turbulent region (see section~\ref{simulations}), and it precedes the splitting of the granule. The third examined granule (last plot in the second row) shows an enhanced power during the splitting process of the large exploding granule. The power is located within the exploding granule in the lower right region, which is where the new downflow lane forms. In this case, the intensity fluctuations of the high photospheric Na I $5896\,$\AA~were analysed because Ca II $8542\,$\AA~data were not available for that day. We also performed an analysis of the velocity fluctuations of the spectral lines for all three exploding granule examples (last row of Fig.~\ref{fig:wavelet}). The first panel in the last row shows extended patches of enhanced wavelet power in more than half of the filed of view. The power is detected earlier than the intensity oscillation enhancement. The cyan extension of the power patches ($[x=3,y=4],[x=3,y=5]$) indicates the areas that are about to show an enhanced power in the intensity oscillation power maps and are along the path of the newly forming downflow lanes. For the third example (last panel in the last row) we detect an island of wavelet power ($[x=3.8,y=2.8]$) in velocity fluctuations at the new downflow lane, which already showed wavelet power for the intensity fluctuations. \\
From the examples we studied, we therefore obtain areas of enhanced wavelet power in the range of 120\,s to 190\,s at the edges of dark cores in the exploding granules as well as locations in which the granule splits. Enhanced oscillations seen in velocity fluctuations have been reported at intergranular and granule fragmentation sites~\cite[e.g.][]{2010ApJ...723L.134B}. We are now able to also support the findings of \cite{Rutten2008} of one exploding granule. They reported intensity and H$\alpha$ velocity oscillations associated with the dark core of the exploding granule.

\section{Summary and conclusions}
We analysed data sets from two instruments, IBIS and IMaX, which had different spatial resolution and cadence as well as noise levels. In addition, we studied \textsc{M\tiny{ANCHA}}3D simulations for a qualitative comparison to our observational findings. For our exploding granule cases, we find several trends and similarities that we summarise below. 

(1) The lifetime of the exploding granules we analysed (10 to 15 minutes) is considerably longer than that of ordinary granules, which is 5 to 9 minutes \citep{1978A&A....62..311M,1999ApJ...515..441H}. In the sample of exploding granules in the IBIS data set, we find evidence for a relation of maximum size, expansion velocity, and lifetime. We find that within the subset of exploding granules, the granules that grow faster reach a larger size and have a shorter lifetime.

(2) While we determined the velocity profiles of the dark core in the photosphere, we found the mean value of the maximum downflow to be 0.24 $\pm$ 0.05\,km/s for the IBIS data set.

(3) There is a trend that exploding granules with diameters smaller than $\sim{3.8}\arcsec$ are more likely to disappear by fading away into a surrounding intergranular lane, whereas larger granules form a new intergranular lane that develops from the dark core. This is true for granules that are larger than or equal to 3.8$\,\arcsec$, with one outlier of one smaller granule with a maximum diameter of 2.6$\,\arcsec$.

(4) We also examined the bisectors extracted at the dark core during the development to an intergranular lane. The downflow velocities at each height of the absorption line are higher during the time when the dark core is formed than the mean value of established intergranular lanes. Whilst the dark core is most pronounced, the velocities in the lower solar photosphere are higher than in the upper photosphere.

(5) Granules that form a new intergranular lane also show a temporal height-dependent shift with respect to the maximum downflow velocity. The time delays for the maximum velocities obtained from the data taken in Fe I 6173.0\,\AA~line are about 48$\,$s and 61$\,$s, while the delays obtained from the data taken in the Fe I 5250.2\,\AA~line are about 20$\,$s and 23$\,$s, respectively. We note this time shift at the point when a downflow is already established throughout the photospheric atmosphere of the downflow lane. This is therefore not to be confused with the time delay caused by the onset of the downflow within the dark core.

(6) In addition, we analysed \textsc{M\tiny{ANCHA}}3D simulations. We studied the change in the behaviour of the velocity evolution at different pixel sample sizes. Similar as in observations, we find a time delay in the times of the maximum downflow when studying the height-dependent downflow velocity of the newly forming downflow lane. This is observed to at least 100 km below the solar surface. The time delay in the maximum velocity is longer for higher resolution (lower binning). A bisector analysis reveals the complex atmosphere and indicates the strength of the gradients in the line-of sight velocity, whereas the study of the effects of pixel binning displays the role of the spatial fine structure of the downflow lane.

(7) A wavelet analysis of the intensity and velocity oscillation in the higher-forming spectral lines revealed an increased wavelet power at the edge of the dark core of exploding granules, as well as during the granule splitting process.

We conclude that the new downflow lane, as observed inside exploding granules, exhibits an intricate fine structure during its formation. This is a horizontal spatial fine structure, which is therefore strongly affected by the spatial resolution of the instrument, and also a vertical fine structure, where the maximum downflow is established at different times throughout the atmosphere. In addition, we presented further evidence that exploding granules are possible candidates for triggering acoustic events.
Further studies of observations with a higher spatial resolution and cadence and a large number of exploding granules are needed to confirm our findings.

\begin{acknowledgements}
Part of the data in this publication were obtained with the facilities of the National Solar Observatory, which is operated by the Association of Universities for Research in Astronomy, Inc. (AURA), under cooperative agreement with the National Science Foundation. 
IBIS has been built by the INAF/Osservatorio Astrofisico di Arcetri with contributions from the Universities of Firenze and Roma Tor Vergata, the National Solar Observatory, and the Italian Ministries of Research (MUR) and Foreign Affairs (MAE). 
This research has made use of the IBIS software reduction package provided and made available by the NSO. This publication makes use of data obtained during DST Service Mode Operations under the proposal ID: SN.2016.4.01. \\

The German contribution to Sunrise is funded by the Bundesministerium für  Wirtschaft  und  Technologie  through  Deutsches  Zentrum  für  Luft-  und Raumfahrt e.V. (DLR), Grant No. 50 OU 0401, and by the Innovationsfond of the President of the Max Planck Society (MPG). The Spanish contribution has been funded by the Spanish MICINN under projects ESP2006-13030-C06 and AYA2009-14105-C06 (including European FEDER funds). The HAO contribution was partly funded through NASA grant number NNX08AH38G.\\

CEF is funded by the SAW-2018-KIS-2-QUEST project and part of this work was carried out within the QUEST project.\\

This work was partially supported by the Spanish Ministry of Science through the project PGC2018-095832-B-I00. It contributes to the deliverable identified in FP7 European Research Council grant agreement ERC-2017-CoG771310-PI2FA for the project ``Partial Ionization: Two-fluid Approach''. The authors thankfully acknowledge the technical expertise and assistance provided by the Spanish Supercomputing Network (Red Espa\~nola de Supercomputaci\'on), as well as the computer resources used: LaPalma Supercomputer, located at the Instituto de Astrof\'isica de Canarias, and MareNostrum based in Barcelona/Spain.\\
ME acknowledges financial support through the SPP1992 under Project DFG RE 1664/17-1.\\
We thank the anonymous referee for the valuable comments and suggestions which have led to the improvement of this paper.
\end{acknowledgements}

\newpage

\bibliographystyle{aa} 
\bibliography{AA2020_38252_ellwarth} 

\begin{appendix}\label{app}
\onecolumn
\section{Additional figures}

        \begin{figure*}[h]
                \begin{minipage}{.5\textwidth}
                        \includegraphics[width=\linewidth]{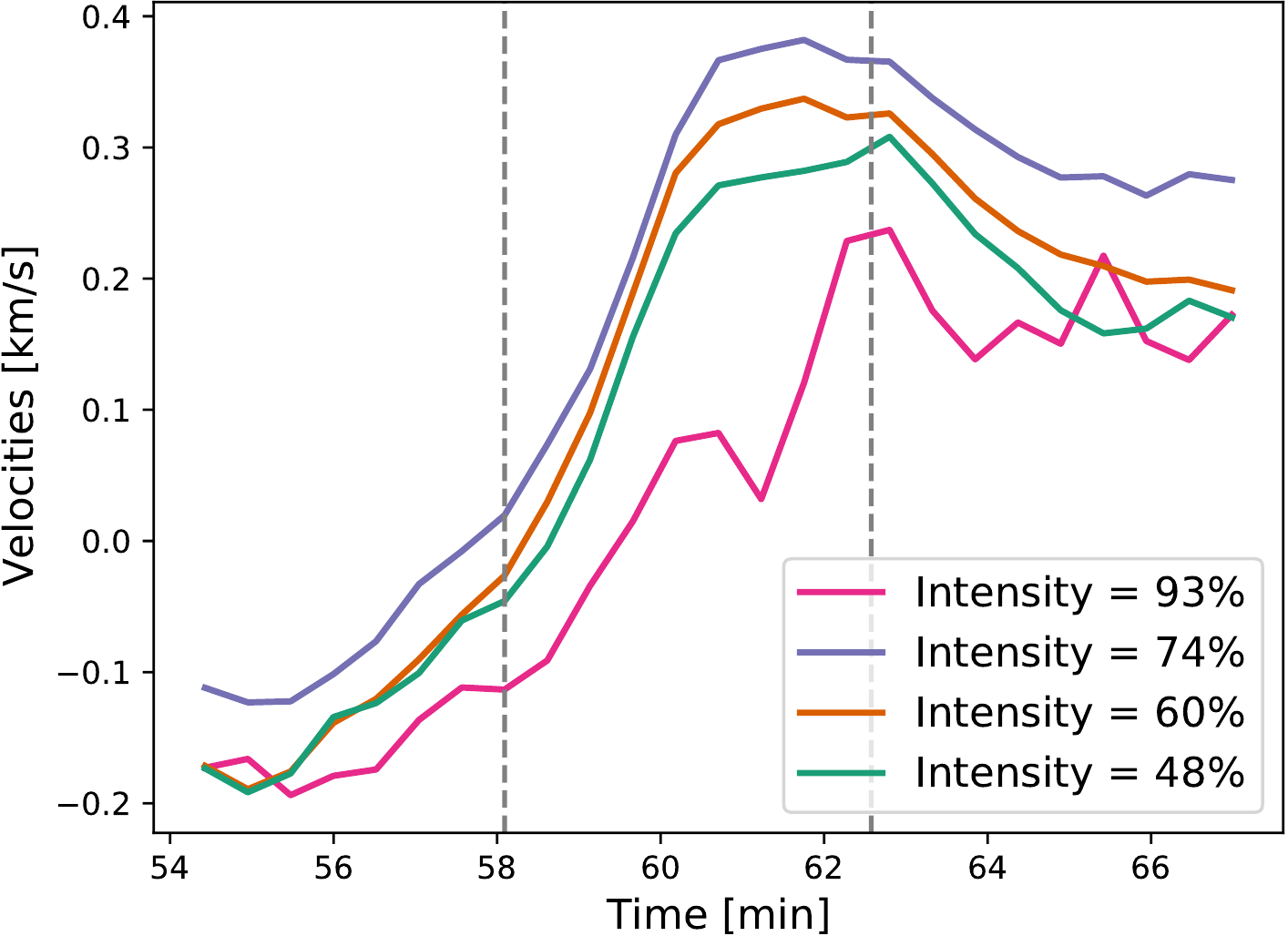}
                \end{minipage}
                \begin{minipage}{.5\textwidth}
                        \includegraphics[width=\linewidth]{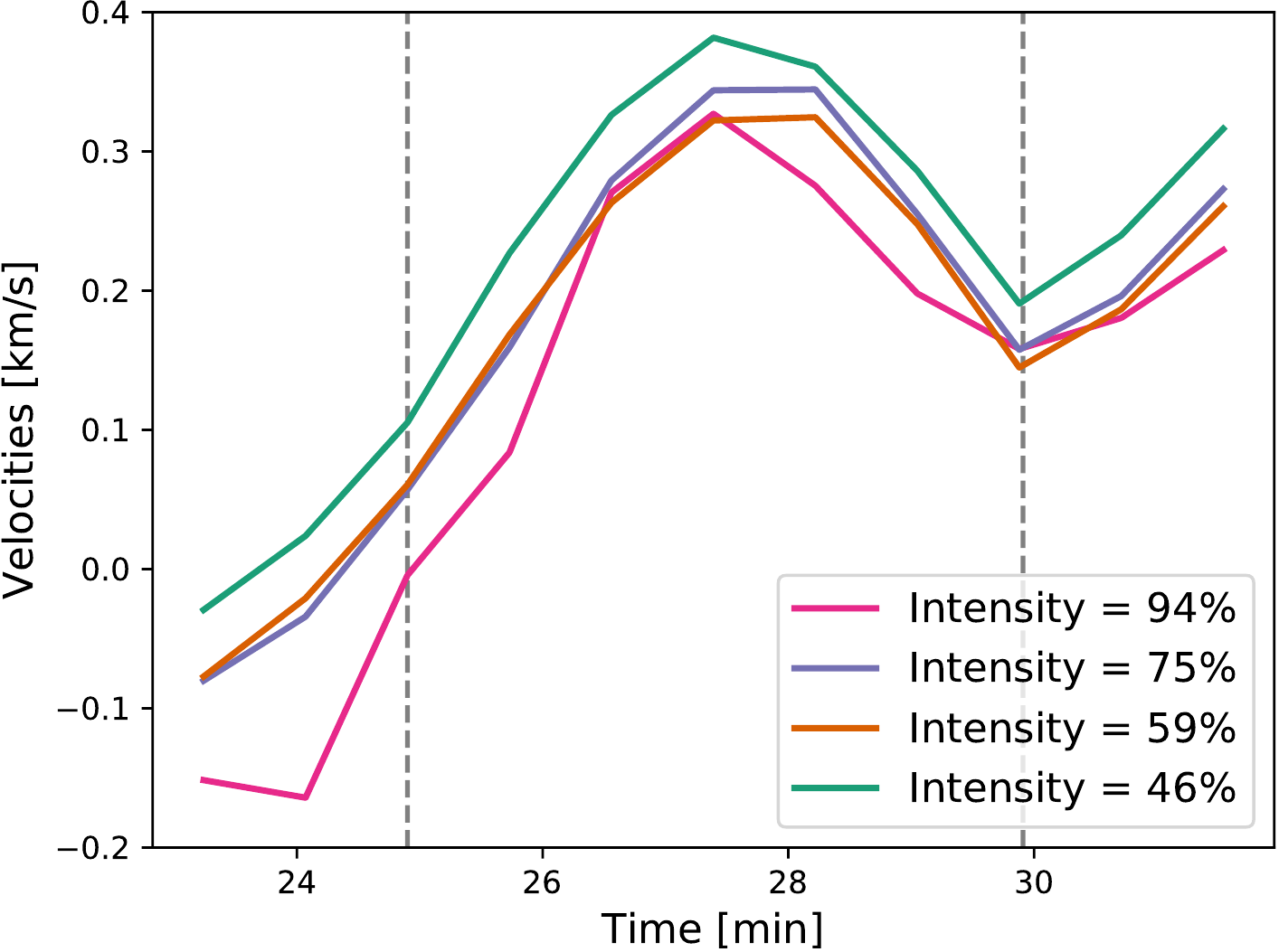}
                \end{minipage}
                \caption{Temporal evolution of bisector velocities calculated from different intensity levels at the location of the dark core. The left plot shows the velocities of granule IBIS 6, and the right plot shows results from granule IBIS 3. The left dashed line marks the time at which the dark core is visible for the first time, and the right dashed line marks the time at which the exploding granule is split into a new generation of granules.}
                \label{fig:bisec_ibis}
                \vspace{0.5cm}
                \begin{minipage}{.5\textwidth}
                        \includegraphics[width=\linewidth]{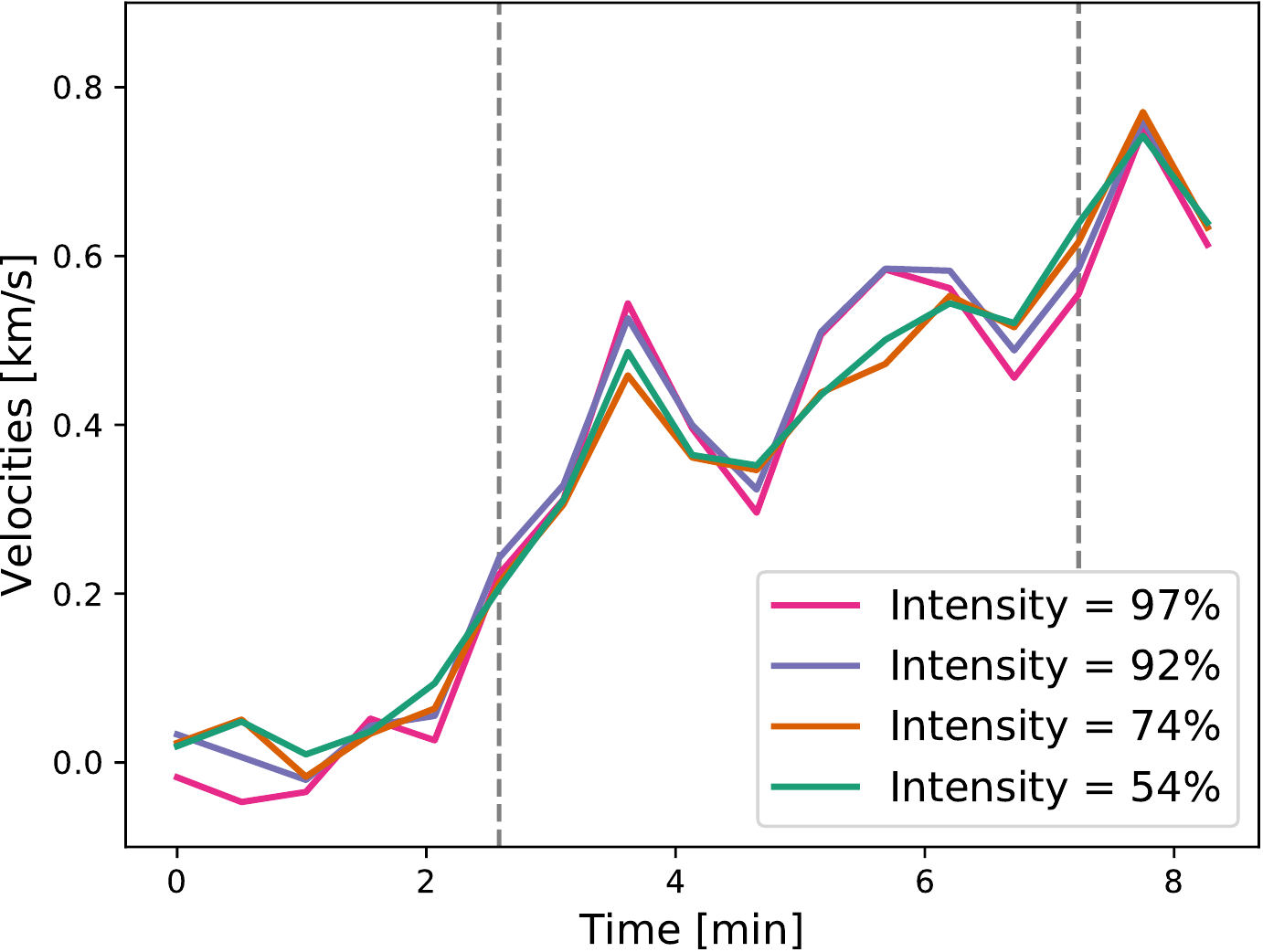}
                \end{minipage}
                \begin{minipage}{.5\textwidth} 
                        \includegraphics[width=\linewidth]{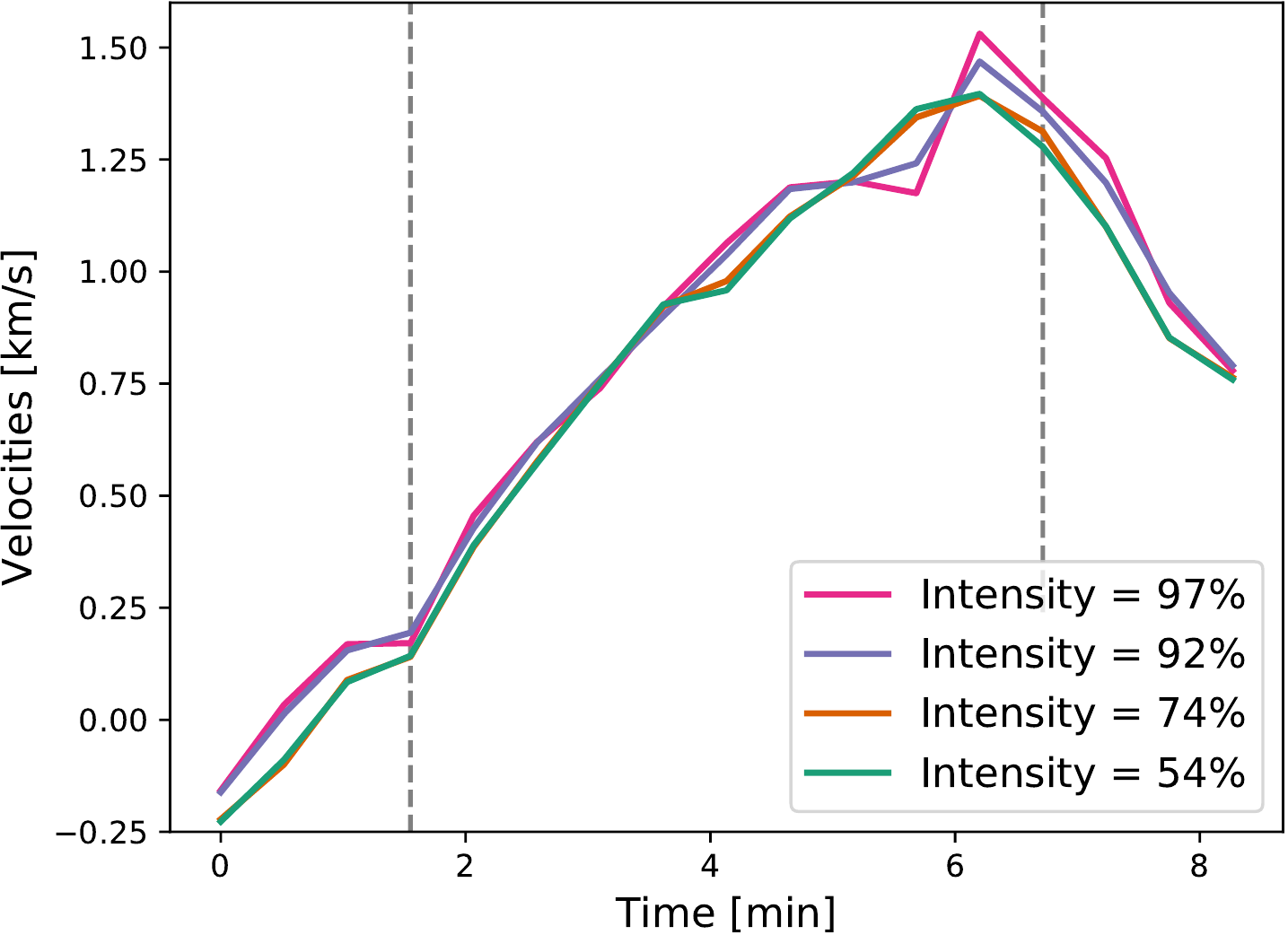}
                \end{minipage}
                \caption{Same as in Fig. \ref{fig:bisec_ibis}, now for IMaX data. The left plot shows the velocities of granule IMaX 1, and the right plot shows results from granule IMaX 2. The left dashed line marks the time at which the dark core is visible for the first time, and the right dashed line marks the time at which the exploding granule is split into a new generation of granules.}
                \label{fig:bisec_imax}
        \end{figure*}

\end{appendix}

\end{document}